\documentclass[journal]{vgtc}

\onlineid{0}

\newcommand{\NAME}{EmbryoProfiler\xspace}

\newcommand{\para}[1]{\vspace{0mm}\noindent\textbf{#1}}

\usepackage[svgnames]{xcolor}
\usepackage{xspace}
\usepackage{tcolorbox}
\definecolor{pink}{HTML}{BC5792}
\definecolor{almond}{HTML}{786B40}
\definecolor{purple}{HTML}{6481C5}

\tcbset{on line, boxsep=1pt, left=0pt,right=0pt,top=0pt,bottom=0pt}

\newcommand{\ACTIVATE}{\tcbox[colframe=pink, colback=pink]{\textcolor{white}{activate}}}
\newcommand{\CHOOSE}{\tcbox[colframe=almond, colback=almond]{\textcolor{white}{choose}}}
\newcommand{\CREATE}{\tcbox[colframe=purple, colback=purple]{\textcolor{white}{create}}}

\vgtccategory{Research}

\vgtcpapertype{application/design study}

\title{\NAME: A Visual Clinical Decision Support System for IVF}

\author{%
  \authororcid{Johannes Knittel}{0000-0002-4889-5232},
  \authororcid{Simon Warchol}{0000-0001-9067-6888},
  Jakob Troidl,
  \authororcid{Camelia D. Brumar}{0000-0002-7924-634X},
  Helen Yu Yang,
  \authororcid{Eric Mörth}{0000-0003-1625-0146}, \\
  \authororcid{Robert Krüger}{0000-0002-6468-8356},
  Daniel Needleman,
  Dalit Ben-Yosef, and
  \authororcid{Hanspeter Pfister}{0000-0002-3620-2582}
}

\authorfooter{
  \item
  	J. Knittel, S. Warchol, J. Troidl, C.D. Brumar, E. Mörth, H. Yang, D. Needleman, and H. Pfister are with Harvard University, C.D. Brumar is also with Tufts University.
  	E-mail: \{jknittel, simonwarchol, jtroidl, cbrumar, ericmoerth, yuy068, pfister\}@g.harvard.edu, dan.needleman@gmail.com
  \item
  	R. Krüger is with New York University (NYU).
  	E-mail: rk4815@nyu.edu.
  \item D. Ben-Yosef is with Tel-Aviv Sourasky Medical Center and Tel Aviv University.
  	E-mail: dalitb@tlvmc.gov.il.
}

\abstract{%
In-vitro fertilization (IVF) has become standard practice to address infertility, which affects more than one in ten couples in the US. However, current protocols yield relatively low success rates of about 20\% per treatment cycle. A critical but complex and time-consuming step is the grading and selection of embryos for implantation. Although incubators with time-lapse microscopy have enabled computational analysis of embryo development, existing automated approaches either require extensive manual annotations or use opaque deep learning models that are hard for clinicians to validate and trust.
We present \emph{\NAME}, a visual analytics system collaboratively developed with embryologists, biologists, and machine learning researchers to support clinicians in visually assessing embryo viability from time-lapse microscopy imagery.
Our system incorporates a deep learning pipeline that automatically annotates microscopy images and extracts clinically interpretable features relevant for embryo grading.
Our contributions include: (1) a semi-automatic, visualization-based workflow that guides clinicians through fertilization assessment, developmental timing evaluation, morphological inspection, and comparative analysis of embryos; (2) innovative interactive visualizations, such as cell-shape plots, designed to facilitate efficient analysis of morphological and developmental characteristics; and (3) an integrated, explainable machine learning classifier offering transparent, clinically-informed embryo viability scoring to predict live birth outcomes.
Quantitative evaluation of our classifier and qualitative case studies conducted with practitioners demonstrate that \emph{\NAME} enables clinicians to make better-informed embryo selection decisions, potentially leading to improved clinical outcomes in IVF treatments.

}


\teaser{
  \centering
  \includegraphics[width=\linewidth, alt={The Timeline View of our EmbryoProfiler system.}]{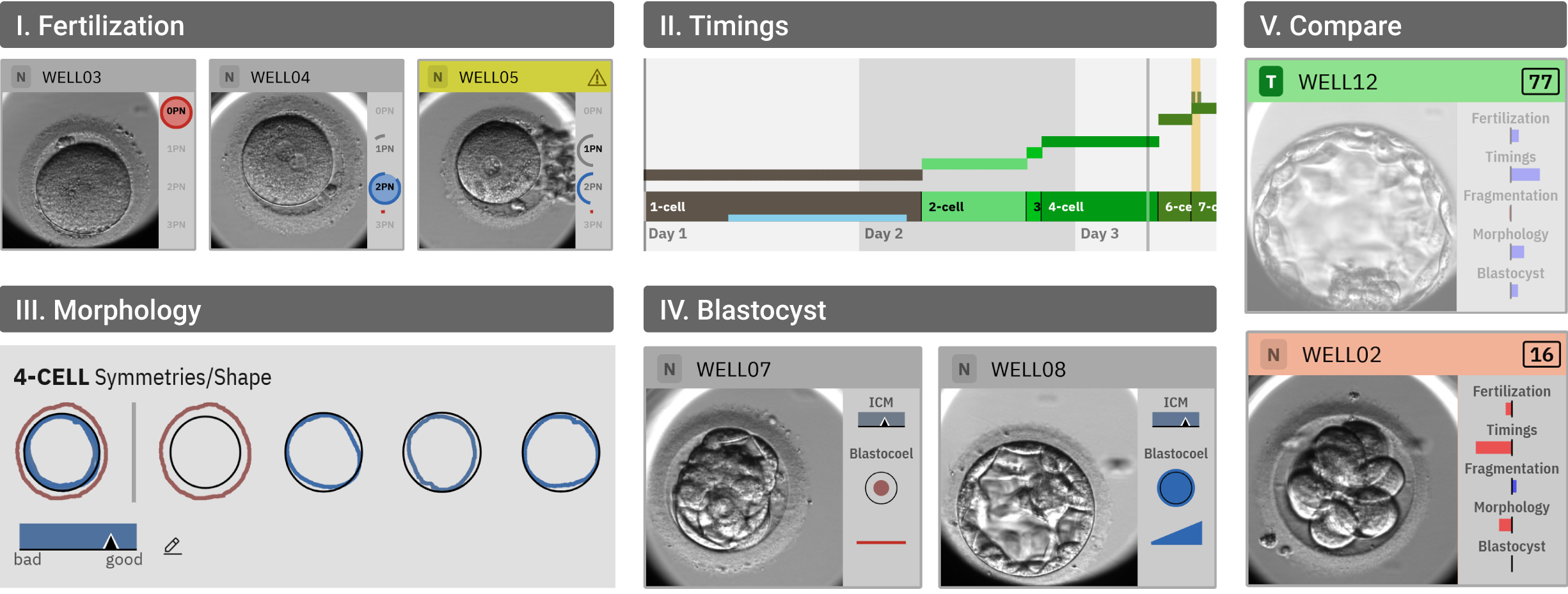}
  \caption{%
  	\textbf{\NAME} supports embryologists in the visual analysis of time-lapse microscopy images for embryo selection through five consecutive dedicated views: (I) assessing fertilization status, (II) verifying developmental stage timings, (III) examining cell morphology and fragmentation, (IV) evaluating blastocyst quality, and (V) comparing embryos with an explainable classifier.
  }
  \label{fig:teaser}
}

\graphicspath{{figs/}{figures/}{pictures/}{images/}{./}}

\usepackage{mathptmx}

\begin{document}

\firstsection{Introduction}

\maketitle

Assisted reproductive technologies (ART), such as in vitro fertilization (IVF) and intracytoplasmic sperm injection (ICSI), represent a crucial development in reproductive medicine, providing hope for individuals and couples facing infertility—an issue affecting approximately 10\% of the US population~\cite{chandraInfertilityImpairedFecundity2013,chandra2013infertility}. Its adoption continues to expand, evidenced by ART contributing to 2.3\% of all US births in 2021, with a steadily increasing trajectory~\cite{centersfordiseasecontrolandprevention2021AssistedReproductive2023}. Nevertheless, the success rate of a single ART cycle remains relatively modest, at approximately 20\% as of 2021, and notably declines with increasing maternal age~\cite{centersfordiseasecontrolandprevention2021AssistedReproductive2023}. Moreover, ART protocols impose substantial financial costs, in addition to being emotionally stressful and physically demanding for patients~\cite{verbergWhyCouplesDropout2008}. Given these challenges, there is an urgent need for innovations capable of enhancing ART outcomes.

A key step in IVF and ICSI procedures involves fertilizing eggs and evaluating the resulting embryos within laboratory settings prior to transfer. For simplicity, we will use IVF as an umbrella term encompassing both IVF and ICSI.
Clinicians use embryo grading—comprising assessments of developmental timings and morphological characteristics—to determine embryos with the highest likelihood of successful implantation and live birth~\cite{nasiriOverviewAvailableMethods2015}.
Ideally, they pick a single embryo for transfer to avoid possibly multiple pregnancies.
Manual grading is inherently tedious, time-consuming, and subject to significant variability across embryologists~\cite{baczkowskiMethodsEmbryoScoring2004,nasiriOverviewAvailableMethods2015}. Recent advances in incubators integrated with time-lapse microscopy, such as the EmbryoScope system~\cite{rubioClinicalValidationEmbryo2014}, present promising opportunities to employ computational techniques for embryo analysis. However, existing computational approaches either rely on extensive manual annotations or utilize opaque end-to-end machine learning models whose predictions are challenging for clinicians to interpret, validate, and trust~\cite{afnanInterpretableNotBlackbox2021}. To truly enhance clinical decision-making, interactive and interpretable systems that effectively support, rather than replace, embryologists’ expert judgment are critically needed.

In this interdisciplinary collaboration among biologists, embryologists, computer vision experts, and visualization researchers from Harvard University and Tel Aviv Sourasky Medical Center, we introduce a visual clinical decision support system called \emph{\NAME}. Designed specifically to assist clinicians analyzing embryo viability from time-lapse microscopy images, \NAME employs a deep learning-based pipeline to visually annotate embryo morphology, highlight critical developmental milestones, and provide transparent viability scores correlated with successful clinical outcomes.
The system's design aligns closely with an interdisciplinary-defined workflow for embryo assessment. It guides embryologists step-by-step through a structured ML-assisted and visualization-driven procedure for verifying fertilization status, pinpointing embryo developmental stages, inspecting morphological features, and monitoring blastocyst maturation. It also enables intuitive side-by-side comparisons among embryos.

Dedicated views support every workflow step, employing small multiples with clear visual cues to facilitate rapid evaluation of critical features. For detailed insights, embryologists can utilize context-sensitive visualizations or interactively explore embryos frame-by-frame via a timeline interface. Throughout, visual indicators explicitly highlight morphological and developmental features linked positively or negatively to clinical success, and allow manual refinement of automatically generated assessments. This clear visual feedback aims to foster user trust in automated evaluation methods.
Ultimately, \NAME is intended to enhance clinical embryo selection for transfer or cryopreservation, potentially improving ART efficacy and patient outcomes.

This paper makes three key contributions toward improving clinical IVF outcomes.
(1) We introduce a semi-automatic, visualization-based embryo grading workflow designed through close interdisciplinary collaboration with clinical embryologists, visualization specialists, and computer vision researchers.
This workflow systematically guides clinicians in assessing embryo fertilization, precise developmental timings, morphological characteristics, blastocyst maturation, and comparative embryo quality analyses.
(2) We further present \emph{\NAME}, a visual analytics system that operationalizes this workflow. It provides embryologists with interactive, intuitive visualization techniques for efficiently examining time-lapse embryo microscopy data, highlighting relevant developmental indicators, and verifying grading assessments.
For each workflow step, we introduce visualizations tailored to the respective tasks, including novel cell shape plots that enable intuitive assessment of cell morphology and symmetry.
(3) Our explainable classifier integrates within our system, leveraging automatically extracted human-interpretable embryo features to estimate embryo viability and predict the likelihood of attaining a live birth upon transfer.
Finally, we present results from an evaluation case study, conducted together with clinical practitioners, in which our visual analytics approach was applied to support embryo selection decisions.
Together, these contributions foster improved interpretability, efficiency, and utility when selecting embryos for IVF treatments.

\section{Related Work}

Recent advancements in automated embryo grading promise improved accuracy and consistency in embryo selection. Complementary research areas, including visual analytics for clinical decision-making and live-cell visualization, provide valuable strategies for deriving insights from complex imaging data. We situate our approach within these related domains and address existing research gaps in the field.

\subsection{Automatic Embryo Grading}
While many embryo grading methods require manual annotations~\cite{bormannConsistencyObjectivityAutomated2020,chavez-badiolaPredictingPregnancyTest2020}, recent automated strategies have leveraged machine-learning algorithms to evaluate specific components of the embryo selection process. These automated methods include assessing sperm quality and determining fertilization status~\cite{zaninovicArtificialIntelligenceHuman2020}.
The training objectives for current systems using time-lapse microscopy imaging (TMI) data without manual annotations encompass a broad spectrum; some concern the development of the embryo into a blastocyst, while others focus on the likelihood of implantation~\cite{bormannPerformanceDeepLearning2020,silver2020datadriven} or chemical pregnancy~\cite{chavez-badiolaPredictingPregnancyTest2020}. Others use the evidence of a fetal heartbeat as a training label~\cite{tranDeepLearningPredictive2019,vermilyeaDevelopmentArtificialIntelligencebased2020,theilgaardlassenDevelopmentValidationDeep2023}. Few approaches, however, attempt to predict live birth, as it involves a number of confounding variables, such as the mother's age~\cite{yangBlastAssistDeepLearning2024}. Many existing methods cannot easily be applied directly for ranking embryos since they either assume a certain developmental stage~\cite{bormannPerformanceDeepLearning2020}, need data for five days of incubation~\cite{vermilyeaDevelopmentArtificialIntelligencebased2020,tranDeepLearningPredictive2019}, or focus on analyzing transferred embryos~\cite{silver2020datadriven}.
The commercial iDAScore grading system~\cite{theilgaardlassenDevelopmentValidationDeep2023}, trained on over $180,000$ embryos from multiple centers using a 3D-CNN model to predict fetal heartbeat pregnancy outcomes, achieved reported AUC scores ranging from $0.62$ (early day-2 transfers) to $0.71$ (day-5 transfers), highlighting the ongoing challenge of accurately predicting clinical embryo viability.

This work builds upon our prior work on deep learning-based computer vision models for detecting and segmenting blastomeres and pronuclei, predicting the level of blastomere fragmentation, and segmenting the zona pellucida and ovum~\cite{jangLearningVectorQuantized2023,leahyAutomatedMeasurementsKey2020}, as well as predicting developmental timings~\cite{lukyanenkoDevelopmentalStageClassification2021}. The derivation of interpretable features is inspired by our previous work BlastAssist~\cite{yangBlastAssistDeepLearning2024}, which computes a subset of interpretable grading features from computer vision models and compares them to human annotations as well as clinical outcomes. We build on and extend these approaches, and propose a novel system for the interactive analysis of time-lapse imaging data, powered by computer vision models and a novel classifier for predicting embryo viability for live birth based on interpretable features.

\subsection{Video Summarization}
Time-lapse microscopy produces vast video data, requiring efficient summarization
as viewing long videos is time-consuming and often not feasible within a clinical context. A recent survey~\cite{afzal2023visualization} categorizes video visualization approaches into video search, video annotation, and video content summarization approaches~\cite{nie2012compact}, among others.
A common strategy for summarizing videos is to extract a small subset of diverse yet representative keyframes~\cite{komlodiKeyFramePreview1998,mundurKeyframebasedVideoSummarization2006,maKeyframeExtractionLaparoscopic2021}, for instance, using clustering methods.  Our approach uses a dedicated stage classifier model to detect boundaries between important events, hence identifying such representative keyframes. Timeline-based visualizations are frequently used for analyzing time series data, particularly in a clinical context~\cite{buiTimeLineVisualizingIntegrated2007,ledesmaHealthTimelineInsightbased2019}. In \NAME, users can hover over the timeline to see the imaged embryo at a precise moment in its development, inspired by preview-based video navigation~\cite{divakaranAugmentingFastforwardRewind2005}.

\subsection{VA for Clinical Decision Making}
Recent research has shown that incorporating pure ML-based recommendations does not necessarily improve the accuracy of clinical decision-making tasks and may even decrease accuracy under certain conditions~\cite{jacobsHowMachinelearningRecommendations2021}. Thus, VA can help contextualize ML-based predictions and help clinicians make informed and data-based decisions about a patient's treatment plans and risks~\cite{informatics8010012}.
Zajac et al.~\cite{zajacClinicianFacingAIWild2023} summarize the opportunities and challenges of introducing AI-based systems into clinical practice, emphasizing the need for close collaborations with practitioners. State-of-the-art VA tools focus on improving the interpretability of ML-based clinical predictions~\cite{jin2020carepre,wentzelExplainableSpatialClustering2020} and efficient visual data summarization~\cite{gotz2014decisionflow, gotz2011visual,zhang2018idmvis,floricelRosesHaveThorns2024}. Highlighting the importance of features on specific predictions is a common strategy to make models more interpretable~\cite{guptaObesityPredictionEHR2022,chengVBridgeConnectingDots2022}, which we adopt as well. \NAME uses machine learning to support the decision-making process by (1) augmenting and summarizing the data for easier manual inspections, as well as (2) providing a score based on interpretable features.

\subsection{VA for Live-Cell Visualization}
VA is commonly used to analyze biological tissue with the guidance of a scientist or physician, for example, in cancer histopathology~\cite{kellerVitessceIntegrativeVisualization2021, jessup2021scope2screen}, radiotherapy planning~\cite{schlachter2019state}, and neuroscience~\cite{beyer2007high, troidl2023vimo}. Cell-Profiler~\cite{carpenterCellProfilerImageAnalysis2006}, a widely adopted open-source tool, allows users to perform a wide variety of analysis tasks, such as cell phenotyping, on different image data modalities and formats.
Other approaches, like Screenit~\cite{dinklaScreenitVisualAnalysis2017}, are more tailored to specific biomedical applications and data structures but provide more interactive visualizations. Facetto~\cite{kruegerFacettoCombiningUnsupervised2020} and Visinity~\cite{warcholVisinityVisualSpatial2022} leverage machine learning by integrating interactive cell clustering and classification, which is based on cell morphology and spatial neighborhoods present in whole-slide tissue images.

Fewer approaches support analysis of time-varying (live cell imaging) data. Pretorius et al.~\cite{pretoriusSurveyVisualizationLive2017} categorize such approaches into spatial embedding, dimensional reduction, spacetime cubes, temporal plots, lineage diagrams, and aggregate visualizations. A common challenge is extracting and visualizing representative states and events.
Loon~\cite{langeLoonUsingExemplars2022} displays main events and features in cancer development by sorting, grouping, and selecting exemplar cells by growth rate and shape. This enables a more compact and comparable timeline visualization for navigation and overview. Their follow-up work, Aardvark~\cite{langeAardvarkCompositeVisualizations2025}, specifically focuses on the visual encoding of cell division in cancerous tissue. With a focus on visual discovery instead of decision-making, the extraction of division timing is not part of their approach.
Similarly, Fangerau et al. ~\cite{fangerauSimilarityAnalysisCell2012} traces paths of cells with features such as shape and construct trajectories representing the lineage. After clustering and dimensional reduction, they can be interactively explored in 2D/3D plots. Cedilnik et al.~\cite{cedilnikIntegrationInformationVolume2008} integrate information and volume visualization for analysis of cell lineage and gene expression during embryogenesis. However, tracking cells and constructing correct lineages from recorded images in a clinical setting is more challenging due to the limited resolution of images, the lack of volumetric data, and the lack of fluorescent markers. Thus, we use the stage prediction model and inferrred pronuclei timings to mark key events.
Duffy et al.~\cite{duffyGlyphbasedVideoVisualization2015} propose a glyph-based video visualization technique to facilitate analysis of sperm quality, addressing an earlier stage of the IVF process.

Most existing approaches for IVF primarily focus on machine learning methods, resulting in challenges related to interpretability and effective human-AI integration.
To the best of our knowledge, our work presents the first visual analytics system designed specifically to support embryo selection using non-invasive, time-lapse microscopy data gathered from incubators increasingly adopted in clinical practice.

\section{Decision-Making in IVF}
\label{sec:dm-in-ivf}
Our primary objective is to support clinicians in analyzing microscopy video data for informed decisions regarding embryo implantation. Our multidisciplinary team comprises clinicians from Tel Aviv Sourasky Medical Center as well as computational biologists, and machine learning and visualization researchers from Harvard University.
This section briefly introduces the background of IVF treatments and outlines the ML-assisted, human-in-the-loop decision-making workflow identified through our research. Additionally, we discuss domain-specific objectives and design requirements established through ongoing iterative discussions over the course of our multi-year collaboration.

\subsection{IVF Treatments}

IVF and its specialized technique, intracytoplasmic sperm injection (ICSI), provide essential solutions for overcoming infertility.
The treatment typically begins with hormonal stimulation, allowing clinicians to retrieve multiple eggs from the ovaries for fertilization~\cite{elderInVitroFertilization2010}. Following fertilization, embryos are cultured for approximately 3–5 days. Initially passing through early cleavage stages (from the \emph{2-cell} to \emph{9+-cell} stage), embryos then compact into a dense cell cluster known as \emph{morula}. Ideally, they progress further to the \emph{blastocyst} stage, characterized by rapid cellular growth and subsequent embryo hatching. Embryologists assess embryo quality by microscopically evaluating morphology~\cite{gardnerHumanGametesPreimplantation2013}. Clinicians then select and transfer the most viable embryo(s) into the mother's uterus, typically opting for a single-embryo transfer to minimize the risk of multiple pregnancies. If pregnancy does not occur, the cycle may be repeated by transferring previously frozen embryos or by initiating another egg retrieval procedure.
Time-lapse incubators like EmbryoScope~\cite{rubioClinicalValidationEmbryo2014} enable continuous monitoring of embryos.
Each embryo is placed in an individual compartment within a specialized petri dish, which is used to fertilize, culture, and observe the embryo.
An EmbryoScope incubator can monitor up to 12 wells on a slide and usually captures images every twenty minutes across up to seven focal planes with a resolution of 500x500 pixels~\cite{EmbryoScopeUserManual}.
Embryologists then analyze these images, compiled into video sequences, to closely assess developmental progression through the stages and morphological features over time.
In this paper, we are concerned with the process of embryo selection only, which starts after the fertilization phase and ends with the selection of embryos to transfer, freeze, or avoid.

\subsection{Decision-Making Process}
\label{sec:decisionMakingProcess}

\begin{figure*}
    \centering
    \includegraphics[width=1.0\linewidth]{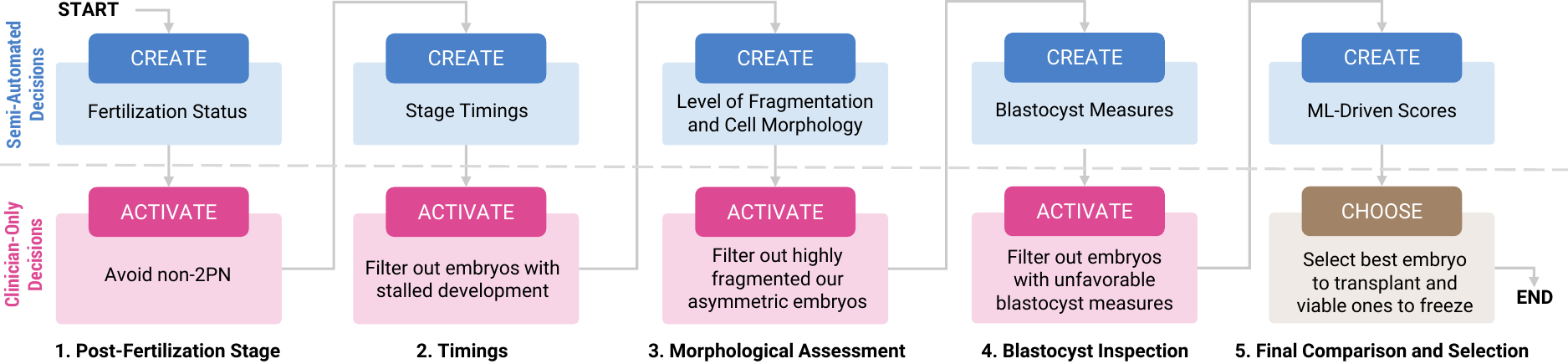}
    \caption{\textbf{Embryo Selection Decision Diagram.} Embryo selection is represented as a decision-making process structured according to the Typology of Decision-Making Tasks~\cite{brumarTypologyDecisionMakingTasks2024}. The sequential steps (1–5) entail semi-automated decisions (blue) and clinician-only decisions (pink and brown).}
    \label{fig:typologyDiagram}
    \vspace*{-0.3cm}
\end{figure*}

Grading embryos for viability encompasses critical decision-making tasks, illustrated in Fig.~\ref{fig:typologyDiagram}.
These decisions affect embryo selection for transfer and can be divided into two types: clinician-only decisions and semi-automatic decisions (i.e., those assisted by machine learning algorithms).
To structure these decision-making processes clearly, we use the \textbf{Typology of Decision-Making Tasks for Visualization}~\cite{brumarTypologyDecisionMakingTasks2024}.
Within this framework, decision tasks fall into three distinct categories: 1) \CREATE: generating, annotating, or synthesizing new information to inform decisions; 2) \ACTIVATE: filtering options based on predefined thresholds or criteria; and 3) \CHOOSE: selecting the best candidates from a set of options.
Applying this taxonomy enables us to systematically decompose the IVF workflow into structured and well-defined decision-making stages. We have identified the following stages:

\smallskip \noindent \textbf{(1) Post-Fertilization Stage.} Clinicians exclusively transfer embryos exhibiting normal fertilization~\cite{reichmanIncidenceDevelopmentZygotes2010}, characterized by exactly two pronuclei (2PN)—one derived from the egg and one from the sperm. Embryologists confirm the automatically assigned embryo categories (0-4PN, 1PN, 2PN, or 3PN) \CREATE \space and subsequently filter for embryos identified as 2PN \ACTIVATE.

\smallskip \noindent \textbf{(2) Timings.} Embryos advance through distinct developmental milestones. The timing of these transitions is critical for embryo grading~\cite{cruzTimingCellDivision2012}. Clinicians validate automatically generated timing annotations of developmental stages \CREATE \space and may choose to exclude embryos exhibiting stalled or delayed progression \ACTIVATE.

\smallskip \noindent \textbf{(3) Morphological Assessment.} Clinicians typically prioritize embryos with minimal fragmentation (cellular debris created during division and not integrated into the blastomeres) and with blastomeres of consistent size and shape~\cite{baczkowskiMethodsEmbryoScoring2004}. Prior research has shown correlations between specific morphological features and successful embryo transfer outcomes~\cite{fancsovitsEarlyPronuclearBreakdown2005,reichmanIncidenceDevelopmentZygotes2010,arakiCorrelationPronucleusSize2018,ezoeCharacteristicsCytoplasmicHalo2021}. Embryologists confirm automatically identified features related to cell morphology, symmetry, and fragmentation levels over time \CREATE. They may subsequently filter out embryos demonstrating significant fragmentation or asymmetry \ACTIVATE.

\smallskip \noindent \textbf{(4) Blastocyst Inspection}. At the blastocyst stage, embryologists employ Gardner scoring~\cite{gardnerBlastocystScoreAffects2000} to evaluate the inner cell mass (ICM)—the cluster of cells destined to become the developing fetus, the fluid-filled blastocoel cavity, and the trophectoderm (outer cell layer)~\CREATE. Embryos with insufficient or suboptimal development may subsequently be filtered out \ACTIVATE.

\smallskip \noindent \textbf{(5) Final Comparison and Selection.} Our system computes viability scores based on human-interpretable grading features from the previous steps \CREATE. Clinicians then comparatively evaluate embryos based on these scores to identify optimal candidates for embryo transfer or cryopreservation \CHOOSE. This critical decision directly influences pregnancy success rates and must strategically balance considerations related to maximizing implantation potential and minimizing the risk of multiple pregnancies.

\subsection{Clinical Goals}
\label{sec:clinicalRequirements}
To guide the design of \NAME, we identified five primary goals:

\noindent
\textbf{G1 – Workflow Integration.} The system should integrate into existing clinical workflows, adapting flexibly to established practices to ensure usability and clinical relevance.

\noindent
\textbf{G2 – Efficacy of Embryo Selection.} Clinicians should be supported making better-informed decisions by clearly identifying embryos with the highest likelihood of resulting in successful pregnancies.

\noindent
\textbf{G3 – Insights into Embryo Development.} Clinicians require detailed contextual insights into embryo developmental trajectories and morphological characteristics to facilitate hypothesis generation and identify correlations between embryo features and clinical outcomes.

\noindent
\textbf{G4 – Transparent and Trustworthy Scores.} The approach must ensure transparency, enabling clinicians to clearly understand the factors influencing machine-learning-derived embryo viability scores and correct potential errors to build trust and confidence in automated grading.

\noindent
\textbf{G5 – Efficient and Streamlined Decision-Making.} Given the limited clinical time available, the solution must be intuitive and streamlined, providing rapid access to critical information and insights without sacrificing analytical depth or accuracy.

\subsection{Design Requirements}

Incorporating feedback from our collaborators, we translated these overarching goals into seven actionable design requirements:

\noindent
\textbf{R1 – Workflow-Oriented Interface.} The interface should clearly reflect and support the identified key workflow steps for embryo viability grading, enabling embryologists to efficiently focus on each task (G1).

\noindent
\textbf{R2 – Contextual Data Overview.} The system should present an intuitive overview of embryo development captured by time-lapse microscopy at critical events, with the ability to drill down for detailed exploration of time-series data when needed (G2,G5).

\noindent
\textbf{R3 – Temporal Aggregation.} Dynamic developmental features and event timings must be aggregated and visually represented clearly, allowing quick assessments and comparisons (G3).

\noindent
\textbf{R4 – Visualization of Grading Relevant Features.} Critical morphological and developmental features such as cell symmetry, shape, and developmental progression should be highlighted, directly supporting embryo grading (G2,G3,G5).

\noindent
\textbf{R5 – Visual Explanations.} Feature contributions to viability scores must be transparently visualized, enabling clinicians to quickly verify scoring logic (G4,G5). The classifier must use human-interpretable features to build user trust (G4).

\noindent
\textbf{R6 – Interactive Feature Correction.} Recognizing clinicians may identify errors or inconsistencies in automated annotations, the system must facilitate intuitive corrections to extracted features, dynamically updating viability assessments accordingly (G2,G4).

\noindent
\textbf{R7 – Accessible Usability for Clinicians.} The interface must be accessible and effective for clinicians, particularly catering to users without specialized visualization or data science training (G1). This includes emphasizing qualitative assessments (viable, non-viable) rather than overly technical feature metrics (G5).

Driven by these goals and design requirements, \emph{\NAME} provides a robust, human-in-the-loop visual analytics solution, explicitly designed to assist clinicians by visualizing relevant embryo features and predicting embryo viability to enhance IVF outcomes.

\section{Data Pipeline and Modeling}

\begin{figure*}
    \centering
    \includegraphics[width=1.0\linewidth]{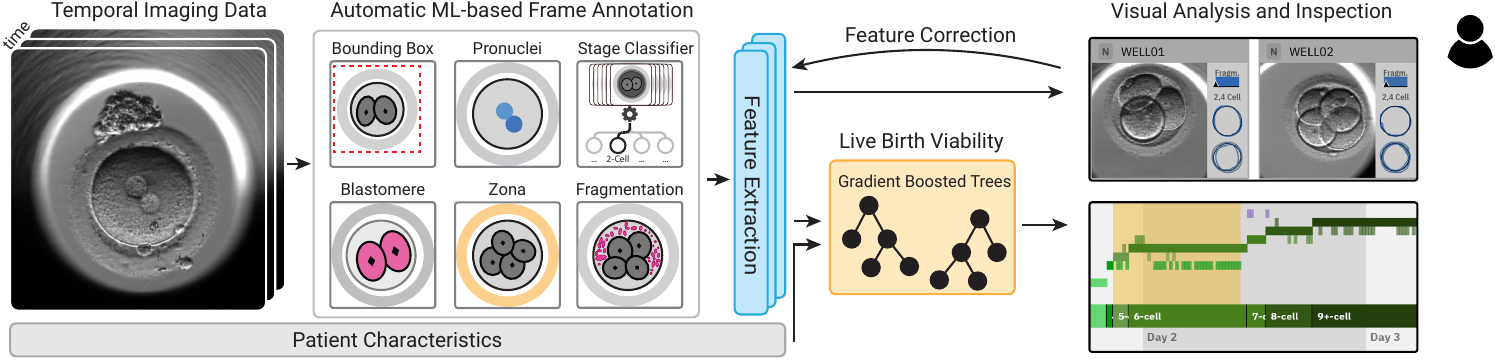}
    \caption{\textbf{System Architecture.}  We first apply computer vision models to annotate each recorded frame. After data cleaning, we extract human-interpretable embryo grading features and combine them with patient characteristics as input to our novel live-birth prediction classifier. Our visual analytics system integrates annotations, derived features, and viability scores into interactive visualizations to support clinical embryo assessment.}
    \label{fig:pipeline}
    \vspace*{-0.3cm}
\end{figure*}

\NAME utilizes microscopy images obtained from EmbryoScope incubators to support embryologists in clinical decision-making.
Each incubator accommodates up to 12 wells (dishes), capturing images every 20 minutes across seven distinct focal planes per well. Our methodology capitalizes on tens of thousands of these time-lapse microscopy images collected throughout each developmental cycle, enriched further by integrating additional pertinent patient characteristics, such as maternal age.
The architecture underlying our approach is illustrated in Fig.~\ref{fig:pipeline}. This section details the system architecture, including a description of our data processing pipeline and the model we trained to compute the viability score assigned to each embryo.

\subsection{Frame-By-Frame Predictions}
\label{sec:cvPipeline}

Building upon our previous work~\cite{jangLearningVectorQuantized2023,leahyAutomatedMeasurementsKey2020}, we employ computer vision models at each timestep to identify and segment cells and pronuclei, measure cell fragmentation levels, segment the zona pellucida from the ovum, and segment the blastocyst into inner cell mass, trophectoderm, and blastocoel.
Additionally, we utilize the stage classifier developed by Lukyanenko et al.~\cite{lukyanenkoDevelopmentalStageClassification2021} to predict the embryonic developmental stage at each frame.
To ensure data quality, we refine these predictions by removing segmentations with low confidence scores, duplicates, and any identified artifacts. Further details regarding our preprocessing methods are provided in App.~B.

\subsection{Feature Engineering}
\label{sec:features}

After data cleaning, we derive human-interpretable features to populate visualizations that clinicians can utilize when evaluating embryo viability for implantation.
These features also serve as input to our viability classifier.
Using these derived characteristics, clinicians can efficiently gain insight into crucial embryo grading criteria, alleviating the need for time-consuming manual annotations and supporting more effective and consistent embryo evaluations (Section~\ref{sec:clinicalRequirements}).
Importantly, our approach contextualizes measurements in a data-driven manner, reducing potential biases in embryo selection strategies by emphasizing features strongly correlated with clinical outcomes.
Our method derives critical timing events, such as pronuclei fading, translates developmental stage predictions into a monotonically increasing time series, estimates pronuclei count, and computes relevant morphological indicators including cell shape and symmetry, average fragmentation levels, and blastocyst growth. A comprehensive list and further details of these derived features can be found in App.~B.

\subsection{Viability Classifier}
\label{sec:classifier}

Our embryo viability classifier serves two main purposes.
First, it communicates a data-driven, automatically derived score that correlates human-interpretable features such as timings and morphological features with the likelihood of live birth.
Optionally, it can integrate specific patient characteristics, such as maternal age and infertility type.
Even though these patient-level attributes remain constant across embryos within the same cycle, they are important in the training process due to their confounding role in predicting live birth outcomes.
Second, the classifier allows us to visualize how individual features correlate with clinical outcomes, guiding embryologists to focus on more influential aspects during embryo assessment.
Importantly, these correlations reflect observational relationships and do not imply causality.

Our dataset comprises $47,336$ embryos with tens of millions of images and corresponding EHR data obtained from the Sourasky Medical Center in Tel Aviv, spanning treatment cycles between 2012 and 2019.
The dataset includes clinician decisions (avoid, freeze/undecided, or transfer) and clinical cycle outcomes (successful live birth).
We divided embryos into training, validation, and test subsets using a 60-20-20 split at the patient level, ensuring embryos from the same patient do not appear across subsets to avoid data leakage.

Given the interpretable features (previous section) and patient characteristics such as age, our goal is to estimate the likelihood of delivering a live birth if said embryo was transferred.
When more embryos were transferred than children born, we apply instance weights to balance the data accordingly.
A central challenge is the unknown outcome for embryos never transferred; we cannot simply assume a negative outcome.
Training exclusively on transferred embryos would introduce significant selection bias, as the model would not encounter non-viable embryos (e.g., highly fragmented embryos). Additionally, many transfers in our dataset occurred at days 2 or 3 rather than day 5.

To address these challenges, we designed a two-step training approach.
First, we trained a maturity classifier to predict whether embryos ultimately developed into expanding blastocysts, reached only morula or early blastocyst stages, or showed no development.
These three maturity categories are strongly correlated with decreasing clinical success and can be objectively assigned without judgment bias.
Subsequently, we leveraged this maturity model to expand and enrich our \emph{training set} for the primary live birth prediction task, using $30,328$ embryos in total.
Specifically, we included frozen and other unlabeled embryos by using maturity-based predictions.
Embryos strongly predicted as non-blastocyst or no development were labeled negative, while embryos strongly predicted as reaching blastocyst stages with expansion were included with a positive label at a reduced instance weight of $0.05$.
To ensure rigorous evaluation, our validation and test datasets comprise only embryos that were explicitly transferred or clearly marked as "avoid".

We use XGBoost~\cite{chenXGBoostScalableTree2016}, an implementation of gradient-boosted trees that have been proven to achieve state-of-the-art performance on classification tasks on such tabular datasets.
Evaluation on the test set demonstrated strong classifier performance, reaching an AUC of $0.83$. Additionally, even in the clinically more challenging scenario restricted solely to embryos previously deemed viable enough for transfer, our approach remains predictive, achieving competitive performance measures of $0.70$ AUC on the test set.
Accurately classifying these embryos presents a greater challenge, reflecting the clinical expertise already invested in their selection.
Building from the finalized model, we computed individual feature correlations with clinical outcomes to support intuitive visualization via color coding, as detailed further in Section~\ref{sec:commonPrinciples}. More details and comprehensive results are provided in App.~C.

\section{Visualization Design}

\begin{figure}[t]
    \centering
    \includegraphics[width=1.0\linewidth]{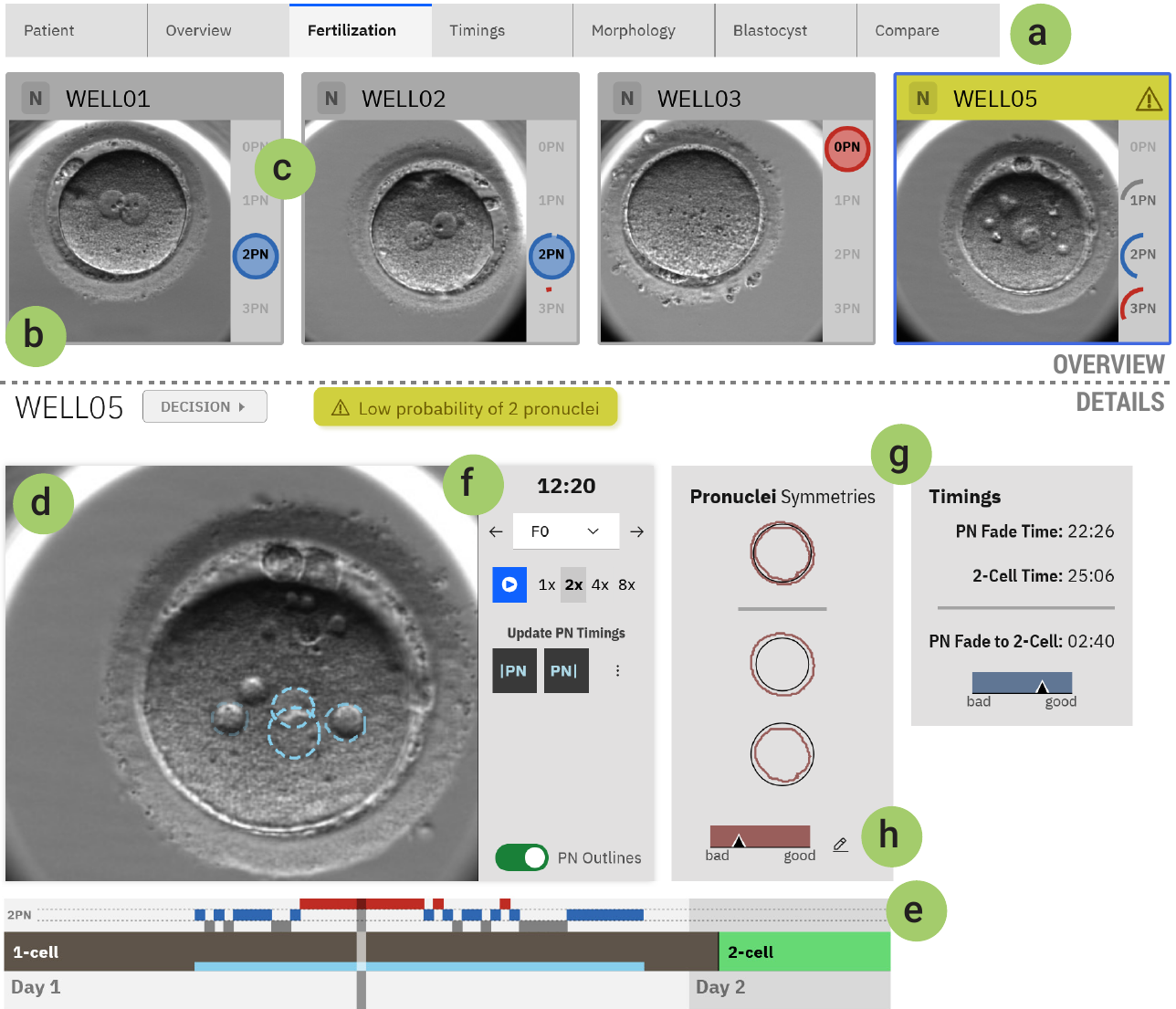}
    \caption{\textbf{Fertilization View.} Tabs guide embryologists through the workflow (a). Each workflow step displays embryos as tiles (b) accompanied by context-dependent visual summaries (c), with instances requiring human verification highlighted in yellow (a). Below, detailed information on a selected embryo is available, including frame-by-frame visualization (d) navigated through an interactive timeline (e). Additional visualizations related to the current task appear above the timeline (e) and adjacent to the viewer (g). Quantitative grading metrics are displayed through an interactive, color-coded slider plot (h), supporting qualitative adjustments.%
    }
    \label{fig:fertilizationView}
     \vspace{-0.3cm}
\end{figure}

\begin{figure*}
    \centering
    \includegraphics[width=1.0\linewidth]{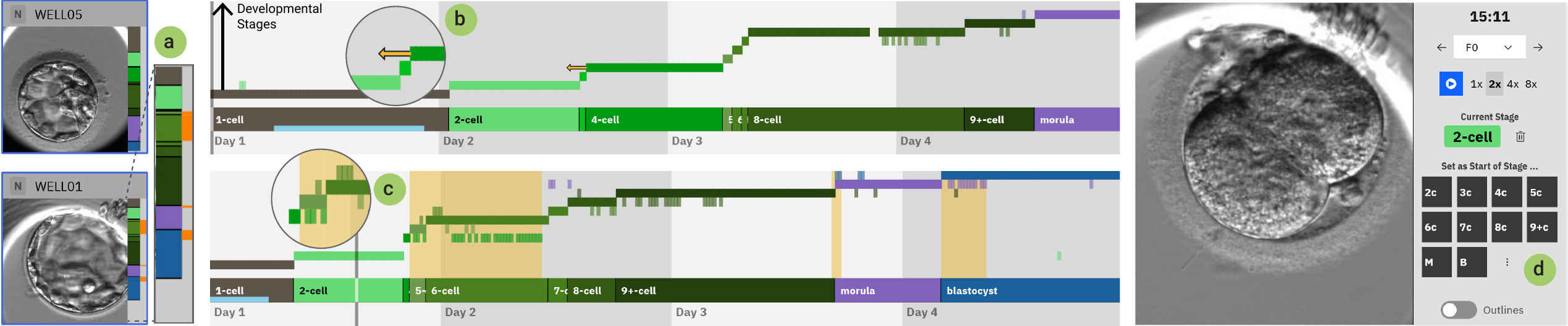}
    \caption{\textbf{Timings View.} Visualization of embryo progression through early cleavage stages up to blastocyst formation using a color-coded timeline (a) and corresponding stage markers (rows above timeline). Alternative predictions that deviate from a strictly monotonic progression are indicated by lighter colors and periods of uncertainty regarding stage timing are highlighted in yellow (c). Arrows (b) point toward the expected timing range when a particular stage starts earlier or later than typical. In this example, both embryos display the characteristic stair-like progression pattern, although automatic annotations show greater uncertainty in the lower embryo. The controls next to the viewer allow embryologists to adjust stage timings (d).}
    \label{fig:timingsView}
    \vspace{-0.3cm}
\end{figure*}

\NAME provides a visual analytics interface designed to support embryologists in the systematic assessment and grading of embryos from time-lapse microscopy images.
The interface is structured into distinct views, corresponding closely to the decision-making process outlined in Section~\ref{sec:decisionMakingProcess} (R1). The workflow starts with an overview display of wells—where users can easily upload and process additional data—
followed by specialized views for checking fertilization status, inspecting developmental timings, assessing morphological criteria such as cell symmetry, rating blastocyst quality, and finally performing side-by-side embryo comparisons supported by our computed viability scores (see Fig.~\ref{fig:teaser}). Additionally, users can input patient-specific characteristics to derive more personalized viability assessments.
This section introduces each view in the order of the workflow.

\subsection{Common Visualization and Interaction Principles}
\label{sec:commonPrinciples}

\noindent\textbf{Small Multiples Visualization.}
We utilize a small multiples approach consistently across all views, displaying each embryo (well) from the current cycle as an individual tile, as shown in Fig.~\ref{fig:fertilizationView}b (R2). Each tile represents the embryo with a carefully selected frame, such as the frame clearly showing the pronuclei for assessing fertilization status. The tile's sidebar contains concise, task-specific summary visualizations (c) that effectively communicate essential grading metrics. Both the chosen frame image and associated summary visuals are customized to support the workflow step under consideration, facilitating quick and informed judgments (R2, R4).
Each tile prominently displays the current decision status (none, transfer, freeze, avoid) in the tile's top-left corner, which users can change through a context menu. For instance, \textit{WELL12} was selected for transfer in Fig.~\ref{fig:teaser}.

\noindent\textbf{Human-in-the-Loop Highlighting.}
To further streamline embryo evaluation, the system highlights embryos with potential data reliability issues using a distinctive yellow header (Fig.~\ref{fig:fertilizationView}a) (R5, R6). This indication alerts embryologists to uncertain or conflicting algorithmic predictions---e.g., ambiguous pronuclei detection---which could negatively influence viability assessments. By explicitly flagging such cases, embryologists can better prioritize their inspection efforts toward ambiguous cases that require further scrutiny.

\noindent\textbf{Timeline-Based In-Depth View.}
Selecting a tile reveals an in-depth view beneath the embryo tiles, comprising two integrated components (R3, R4). The timeline-based frame viewer (Fig.~\ref{fig:fertilizationView}d, e) allows inspection of embryo images across recorded time points, with helpful, context-sensitive interactive controls (f). Users can navigate through frames by hovering over the timeline or replaying the video from the current time point. Our collaborators emphasized the need to quickly scroll through focal planes, so users can easily adjust them via mouse wheel while positioned over the timeline in addition to a dropdown menu. Relevant task-specific temporal features are visualized clearly above the timeline to guide embryologists efficiently toward crucial frames (e).\looseness=-1

\noindent\textbf{Grading Feature Visualizations.}
Alongside the frame viewer, contextualized visualizations present detailed evaluations of embryo grading criteria (Fig.~\ref{fig:fertilizationView}g), offering users complementary insights for accurate assessment. For highlighted embryos, the detail view further lists the reason for the warning (e.g., uncertainty about number of pronuclei).
Across the interface, we consistently employ a red-gray-blue color scheme to intuitively indicate correlations with lower or higher embryo viability (R5, R7).\looseness=-1

\begin{figure*}
    \centering
    \includegraphics[width=1.0\linewidth]{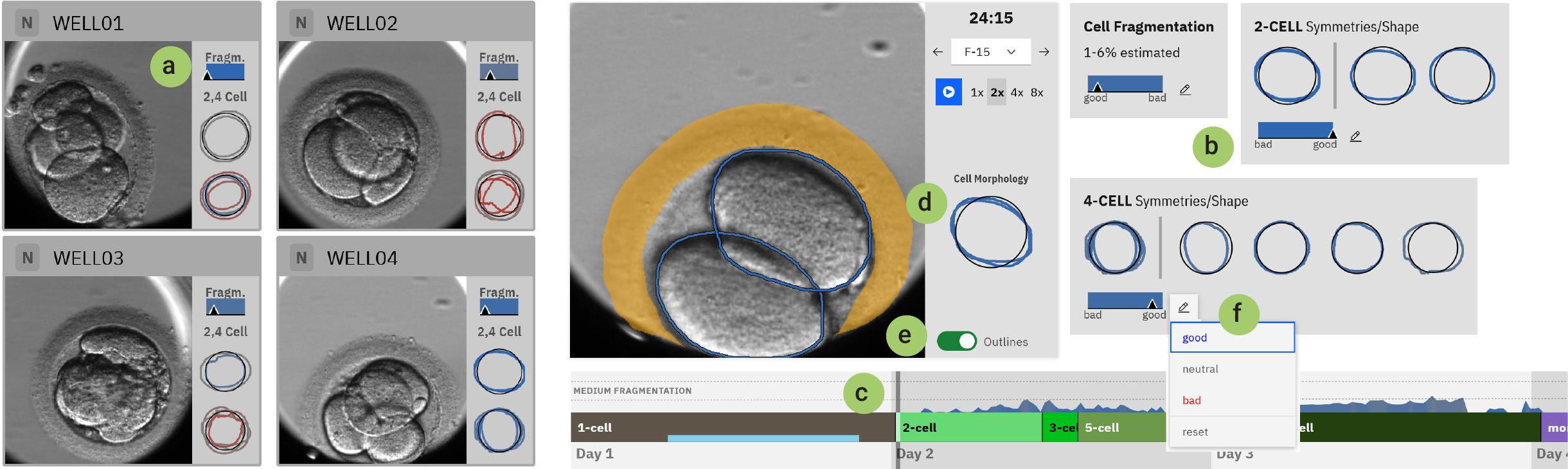}
    \caption{\textbf{Morphology View} enables assessment of cell symmetry, shape, and fragmentation during early cleavage stages. The overview tiles provide concise visual summaries of fragmentation and cell shapes at the 2-cell and 4-cell stages (a), while the detailed view offers expanded visualizations for closer inspection (b) and means to adjust automated assessments (f). Fragmentation levels over time are depicted above the timeline (c). Additionally, the frame viewer dynamically illustrates the cell shapes of the current frame (d), with optional overlays of cell annotations (e).}
    \label{fig:morphologyView}
     \vspace{-0.3cm}
\end{figure*}

\noindent\textbf{Feature Adjustments.}
For quantitative grading criteria, such as fragmentation levels, we introduce intuitive slider-style plots (Fig.~\ref{fig:fertilizationView}h). An arrow visually situates the embryo's feature value along a linear continuum ranging from ``bad'' to ``good''. Additionally, we reinforce this visual encoding by applying a matching background color to provide redundancy, thereby facilitating both rapid, preattentive feature evaluation and more precise visual comparisons (R7).
To accommodate manual adjustments for trustworthy human-AI decision-making (R6), users can directly adjust these feature assessments through an intuitive menu, selecting categories such as good, neutral, or bad (Fig.~\ref{fig:morphologyView}f).

\subsection{Fertilization View}

\begin{figure}
    \centering
    \includegraphics[width=1\linewidth]{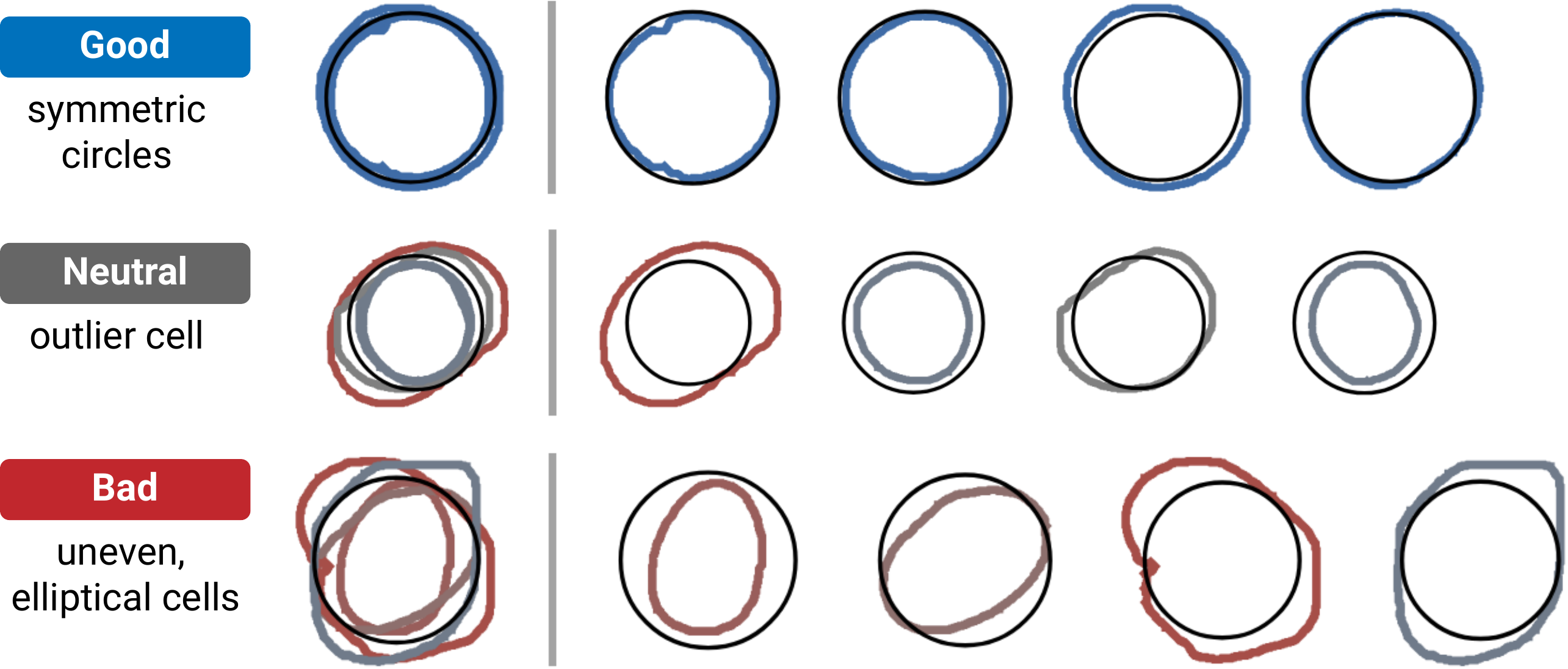}
    \caption{\textbf{Cell Shape Plots} for rapid evaluation of cell morphology and symmetry. \textit{Left:} Centrally aligned, superimposed cell outlines, color-coded by similarity to the ideal circular shape (thin black circle) calculated from the average cell area. \textit{Right:} Extended view showing individual cell outlines side-by-side, each with a reference circle (black) based on the mean area of the remaining cells.%
    }
    \label{fig:cellShapePlots}
    \vspace{-10px}
\end{figure}

The \emph{Fertilization View} (Fig.~\ref{fig:fertilizationView}) is designed to help embryologists confirm normal embryo fertilization. Normally fertilized embryos contain exactly two pronuclei (2PN), which typically disappear shortly before cell division begins. Embryos with zero, one (1PN), or three or more pronuclei (3+PN) are excluded from transfer. While pronuclei count is the main criterion, embryologists also consider features such as pronuclear symmetry, inter-pronuclear distance, and the timing from pronuclear fading to cleavage.

Next to each representative frame (Fig.~\ref{fig:fertilizationView}b), we provide an uncertainty-aware visualization of automated pronuclei counts using circular plots (c). Each plot shows the proportion of frames, within the visibility window, classified into each category. Near-complete circles with filled backgrounds visually denote clear cases: blue for optimal (2PN), red for abnormal (0PN or 3+PN), and gray for the intermediate (1PN) category, which may result from partial occlusion.
These plots also serve as interactive buttons for embryologists to adjust embryo status.
The representative frame shown is the one that best displays pronuclei, providing quick visual confirmation.

The detailed view (Fig.~\ref{fig:fertilizationView} bottom) helps embryologists manage ambiguous cases. Above the timeline, the detected category is displayed for each frame (Fig.~\ref{fig:fertilizationView}e), enabling rapid navigation to problematic frames.
The visualization automatically emphasizes the early developmental stages, typically up to two days.
Additional plots next to the viewer highlight key metrics such as pronuclear symmetry and the interval between pronuclear fading and first cleavage (g). Further discussion of these plots is provided in the \emph{Morphology View} subsection.

\subsection{Timings View}

After fertilization, embryos undergo cleavage, with rapidly dividing cells forming an expanding blastocyst.
Accurate timing of early divisions is crucial for embryo grading.
The \emph{Timings View} shows development across eleven stages using stacked bars (Fig.~\ref{fig:timingsView}a), mapping the eight early stages (2-cell to 9+-cell) onto a light-to-dark green gradient for clear contrast with the later and more mature morula (violet) and blastocyst (blue) stages.

Above the timeline, we visualize progression through the stages using dedicated rows for each stage, in addition to color coding. This creates a stair-step pattern that intuitively depicts the extent and speed of cleavage—slow progression appears as gradual ascents, while rapid or abrupt changes result in larger vertical jumps.
This layout also accommodates conflicting stage predictions, which are displayed as points at different $y$ positions and highlighted with brighter colors (Fig.~\ref{fig:timingsView}c), making discrepancies from the final stage assignment (after monotonic mapping) immediately apparent. A clean stair-step pattern reflects confident stage detection, whereas scattered bright points indicate uncertainty or problematic classifications.

\NAME further assists in identifying data quality issues by: (1) highlighting sequences with alternative predicted timings in yellow (Fig.~\ref{fig:timingsView}c); and (2) marking significant deviations from expected stage onset with yellow arrows (Fig.~b). Stage mapping controls (d) in the frame viewer allow embryologists to refine stage assignments by setting the start frame for a stage or removing assignments as needed.

\subsection{Morphology View}

Embryologists assess cell symmetry and overall cell shape in embryos at early developmental stages, particularly the 2-cell and 4-cell stages.
Consistent with established practice, our data confirm that embryos exhibiting significant cell size (area) variability or reduced roundness have a lower probability of resulting in successful live births.
Moreover, the degree of cellular fragmentation is another critical evaluation criterion; embryos demonstrating extensive fragmentation are avoided, as higher fragmentation correlates with poorer developmental outcomes.

We introduce \emph{cell shape plots} to support embryologists in quickly and objectively evaluating cell shape and symmetry.
In these plots, segmented cell outlines are centrally aligned and superimposed, with each outline color-coded according to how closely it achieves the ideal shape and symmetry.
A thin black circle representing the ideal shape (from the average cell area) serves as a reference.
Thus, with perfectly round and symmetric cells, the plot converges to a clear blue circle overlapping the thin black reference circle, shown at the top left in Fig.~\ref{fig:cellShapePlots}.
Notable deviations in size or shape become immediately apparent (middle left). Greater deviations produce increasingly indistinct edges reminiscent of a fuzzy solar eclipse, while severe shape variations result in a cluttered arrangement of red polygons (bottom left).

Our detailed \emph{Morphology View} extends this by displaying each cell outline side-by-side (Fig.~\ref{fig:morphologyView}b), each with its own black reference circle that represents the ideal shape derived from the average cell area of the other cells.
This additional visualization facilitates rapid identification and clear characterization of shape and size deviations, strongly supporting detailed morphological analysis.
Cell shape plots are computed for both the 2-cell and 4-cell stages, using frames that are most representative of average morphology; the 4-cell stage frame also serves as the overview tile.
As users scroll through frames, the viewer dynamically updates the corresponding cell shape plot (Fig.\ref{fig:morphologyView}d), and users can overlay detected cell outlines on images (e).
The cell shape plot is also used in the \emph{Fertilization View} to assess pronuclear morphology (Fig.\ref{fig:fertilizationView}g).
For fragmentation, we complement a slider plot of average values with an area chart above the timeline showing fragmentation evolution over time, with background colors reflecting the viability-driven fragmentation assessment to help users quickly interpret embryo quality changes (Fig.~\ref{fig:morphologyView}c).

\subsection{Blastocyst View}

\begin{figure}
    \centering
    \includegraphics[width=1.0\linewidth]{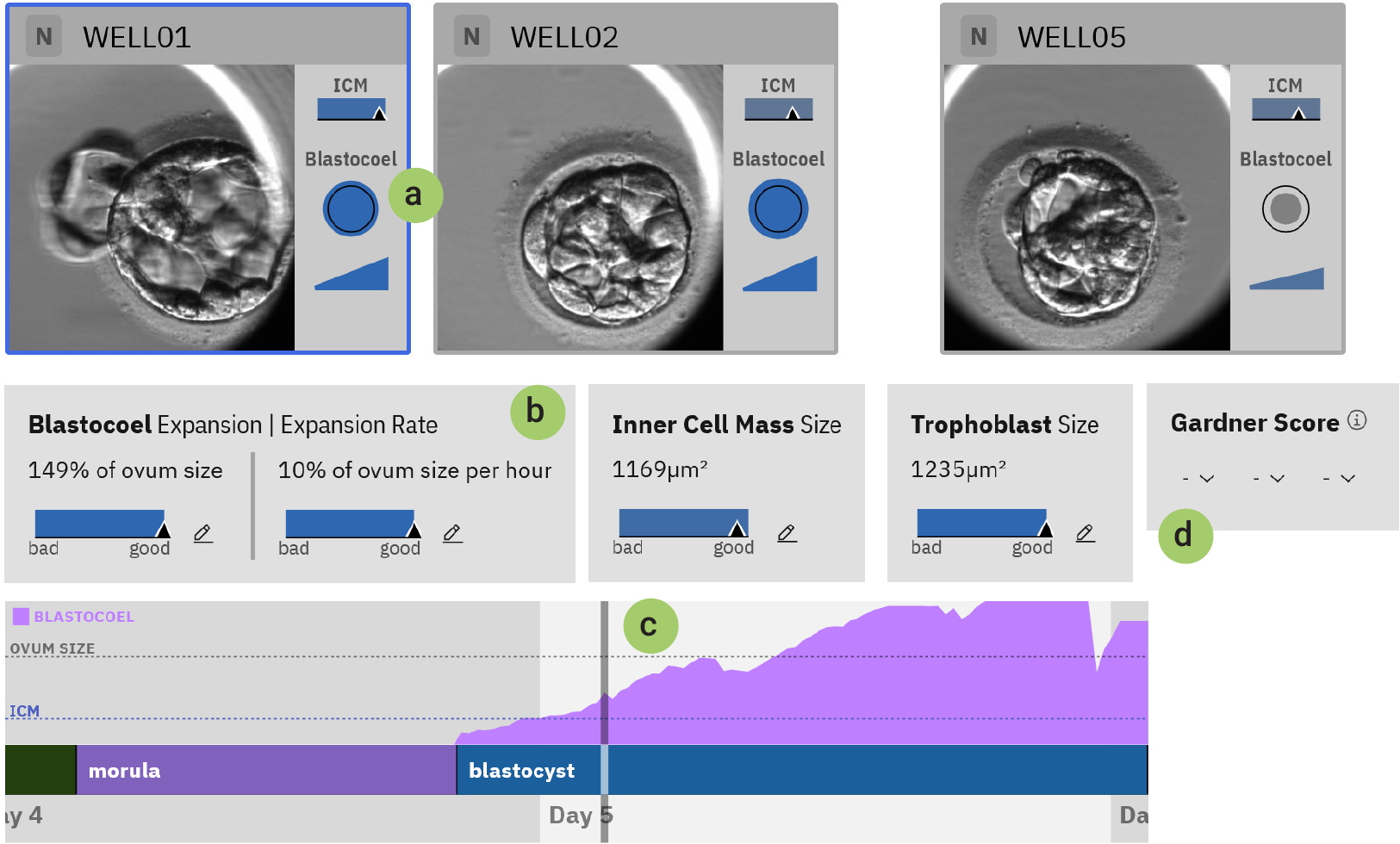}
    \caption{\textbf{Blastocyst View.} The morphology of the inner cell mass (ICM) and the expansion of the blastocoel cavity are key criteria in blastocyst grading. Each tile is accompanied by a color-coded schematic visualization (a), depicting the current blastocoel area (filled circle) relative to the original ovum size (thin black circle outline). The triangle's slope illustrates the rate of cavity expansion. The detailed view provides a time-series plot of blastocoel cavity size (c)}
    \label{fig:blastocystView}
     \vspace{-0.3cm}
\end{figure}

Once embryos reach the blastocyst stage, the blastocoel cavity forms and progressively expands until embryo hatching occurs.
At this stage, embryologists commonly evaluate three components—the inner cell mass (ICM), trophectoderm, and blastocoel expansion~\cite{gardnerBlastocystScoreAffects2000}.
Our summary plots placed to the right of each embryo tile effectively illustrate the ICM area, blastocoel size, and growth rate (Fig.~\ref{fig:blastocystView}a).
Specifically, the maximum observed blastocoel area is depicted as a filled circle, proportional to the original ovum size, which is indicated by a thin black circle.
This schematic visualization helps users assess the relative proportions of the blastocoel cavity within the embryo, aligning directly with the first criterion of the Gardner score.
Additionally, a filled triangle positioned below encodes the expansion rate of the blastocoel, where steeper triangle angles signify faster growth.
Ideally, embryos demonstrate rapid expansion and formation of large cavities.
Consistent application of our color scale further aids embryologists in interpreting these metrics in context.

For embryos selected for in-depth analysis, we also present the progression of blastocoel area over time as a magenta curve positioned above the embryo timeline (Fig.~\ref{fig:blastocystView}c).
Dashed reference lines denote the original ovum area and estimated inner cell mass area, providing essential context.
Embryologists can also rely on a range of additional features (b). For example, they may optionally assign a Gardner score for personal reference (d). Although this score does not impact our automated viability classification, it serves as a familiar bridge linking traditional grading techniques with our comprehensive, machine learning-supported evaluation approach.

\subsection{Compare View}

\begin{figure}
    \centering
    \includegraphics[width=1.0\linewidth]{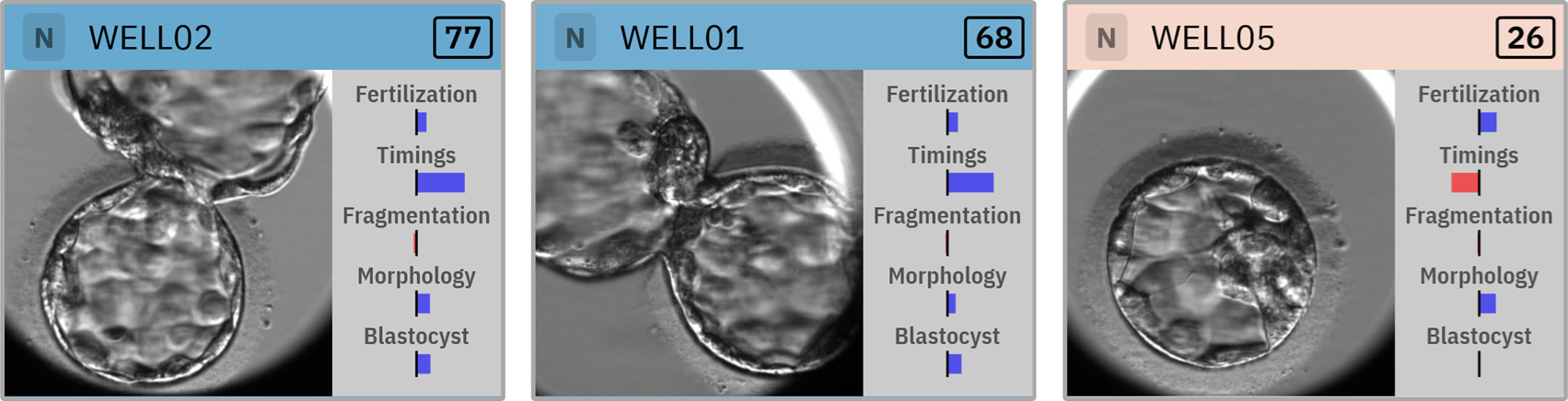}
    \caption{\textbf{Scoring.} The comparison view incorporates our viability score (top right). The diverging plot adjacent to each embryo visualizes feature importance through SHAP values, grouped according to workflow steps. In this example, the first two wells contain viable embryos, whereas cleavage-timing features negatively influence the score of \textit{WELL05}.}
    \label{fig:compareShap}
    \vspace{-0.3cm}
\end{figure}

The previous workflow steps focused primarily on analyzing embryos individually, though our overview visualizations inherently support comparative analyses.
Specifically designed for side-by-side embryo comparisons, the \emph{Compare View} facilitates the final assessment process.
Aligned with our collaborators' core clinical goal of integrating data-driven metrics into embryo grading (G2), this view provides the viability scores generated by our interpretable classifier as numerical values (0–100) at the top-right corner of each embryo (Fig.~\ref{fig:compareShap}). These scores are further visually reinforced by the background color of the header. In contrast to earlier views, embryos here are sorted according to their predicted viability, streamlining comparative decision-making.

To interpret the factors influencing each viability score, we employ SHAP values~\cite{lundbergLocalExplanationsGlobal2020}. Given the additive nature of these values, we aggregate individual features into five major categories: fertilization, timings, fragmentation, morphology, and blastocyst formation.
We illustrate feature impacts through diverging bar plots (G4, R5), where blue bars extending to the right represent features positively influencing the viability score, and red bars extending to the left represent negative influences. However, it is crucial to recognize that instance-level explanations, such as SHAP values, can fluctuate substantially even due to minor changes in a single feature.
Therefore, SHAP values do not inherently represent the absolute quality of individual aspects but rather their relative contribution within a specific context. For example, if one feature strongly indicates a low probability of embryo viability, other potentially negative features might appear to have diminished influence, despite also being unfavorable.
This contextual sensitivity motivates our grouping of features rather than focusing on individual ones.

\begin{figure}
    \centering
    \includegraphics[width=1.0\linewidth]{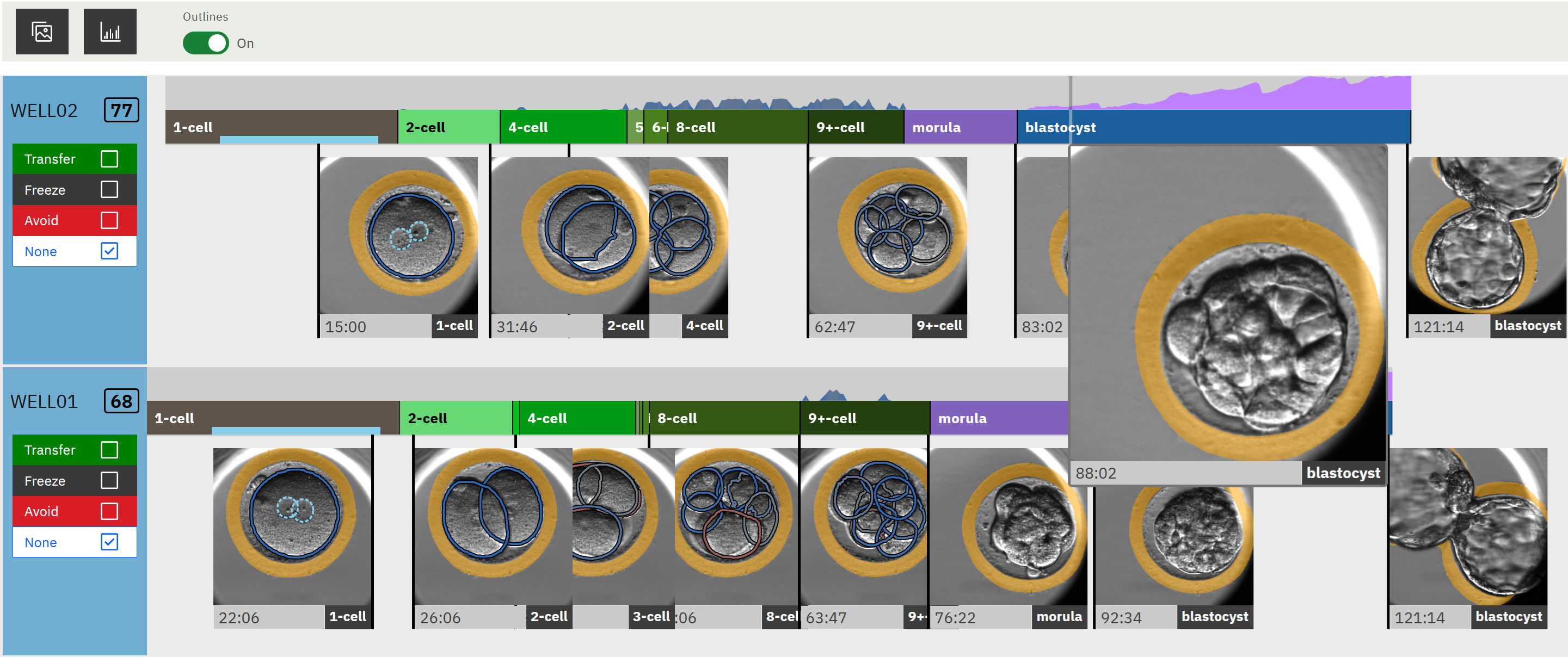}
    \caption{\textbf{Comparison View.} Embryologists can compare multiple embryos side by side using synchronized timelines aligned by pronuclei fading. Key developmental stages are shown with representative frames and optional overlays (cell boundaries, pronuclei, zona pellucida). Charts illustrate fragmentation during early cleavage and blastocoel expansion (magenta) at the blastocyst stage. Frames and charts are optionally hidden for compact viewing. Hovering over the timeline enlarges the corresponding frame.}
    \label{fig:compareRows}
     \vspace{-0.3cm}
\end{figure}

Embryologists can select multiple embryos to be visualized as compact rows below for comparison (Fig.~\ref{fig:compareRows}).
Each row features a familiar timeline accompanied by representative frames illustrating critical developmental milestones.
Hovering over the timeline reveals the corresponding frames through an enlarged tooltip for easy examination.
Incorporating feedback from clinicians, timelines are aligned to pronuclei fading for accurate developmental timing comparisons.
Area charts show cell fragmentation in early stages and, if applicable, blastocoel expansion during the blastocyst stage.
Buttons allow users to toggle the visibility of charts and representative frames (Fig.~\ref{fig:compareRows} top left), enabling flexible focus on timings, fragmentation and blastocyst expansion metrics, morphological evaluations, or all together for comprehensive comparison.

\section{Evaluation}

We demonstrate the effectiveness and usability of \NAME through a case study and a user study with four embryologists.
App.~A describes our iterative design process that led to the current version of our approach.
For the quantitative evaluation of our viability classifier, we refer readers to Section~\ref{sec:classifier} and App.~C.

\subsection{Case Study}

This case study emerged from collaborative Zoom sessions with one computational biologist and one embryologist to discuss the design of the approach in a pair analytics~\cite{arias-hernandezPairAnalyticsCapturing2011} setting where we presented different functionalities of the approach and collected feedback.
The cycle was chosen from a small random subset of the entire dataset of our collaborators (Section~\ref{sec:classifier}) and consisted of four embryos spanning 350 time steps, resulting in $9,800$ images to analyze.
After loading the data, the analyst first evaluated embryo fertilization status using the Fertilization View (Fig.\ref{fig:fertilizationView}). The first two wells clearly exhibited successful fertilization (2PN, shown as filled circles), as confirmed by representative frames. In contrast, the third well appeared non-fertilized, prompting immediate classification of this embryo as one to avoid using the menu at the top left.
The fourth well (in yellow) presented some uncertainty, with frames variably classified as 1PN, 2PN, and 3PN. Upon detailed examination of frames labeled as 3PN (Fig.\ref{fig:fertilizationView}e) and toggling detection outlines, the analyst identified a misclassification: a vacuole had erroneously been identified as a third pronucleus. The embryo was then manually adjusted to a correct classification of 2PN. However, the cell shape plot indicated uneven pronuclei sizes (Fig.~\ref{fig:fertilizationView}h), which the analyst noted as an unfavorable, albeit non-disqualifying feature.

Next, evaluation of developmental timing in the Timings View revealed no major data quality issues but did identify a delayed transition to the 4-cell stage for \textit{WELL05}, indicated by an arrow in the stair-step visualization (Fig.\ref{fig:timingsView}b). Upon closer inspection (by hovering over this event), the delay was confirmed, marking another negative aspect for this embryo. Conversely, in the Morphology View, the cell shape plots suggested overall favorable morphology, depicted as uniform blue circles. The Blastocyst View (Fig.\ref{fig:blastocystView}) further confirmed strong expansion and growth rates for embryos in the first two wells (with \textit{WELL02} slightly outperforming \textit{WELL01}), clearly surpassing \textit{WELL05}.
This finding was supported by the Compare View (Fig.~\ref{fig:compareShap}), where the first two wells scored highly, while \textit{WELL05} was ranked considerably lower.

Given these assessments, the analyst decided to continue monitoring \textit{WELL05} for an additional day prior to determining whether it should be frozen or excluded. The two highest-ranked embryos, \textit{WELL01} and \textit{WELL02}, were selected for an in-depth comparison (Fig.~\ref{fig:compareRows}). Timeline analysis revealed slightly faster development and a more prolonged period of rapid blastocyst expansion for \textit{WELL02}. Moreover, closer inspection of the 4-cell stage keyframe showed superior blastomere symmetry and less cell overlap in \textit{WELL02}. Considering these factors in combination with overall assessment scores, the analyst ultimately selected \textit{WELL02} for immediate transfer, opting to freeze \textit{WELL01}.

\subsection{User Study--Method}

\textbf{Participants:}
We recruited four embryologists from the Tel Aviv Sourasky Medical Center who had no prior exposure to our system (all female). These embryologists typically perform embryo annotation and grading incrementally using the commercial VitroLife system, assessing fertilization status on day 1 and subsequently verifying suggested developmental timings each day until recording completion.

\noindent\textbf{Setup:} Participants initially completed a pre-study survey describing their existing workflow and recorded the typical duration required to annotate embryos. Following this, they were introduced to our approach through a concise, seven-minute instructional video at the beginning of a remote session conducted via Zoom.
During the study session that lasted between 60 and 75 minutes, participants used our hosted web application to grade embryos comprehensively from day 1 through day 5, utilizing their own equipment.
Their primary task was to select the single most viable embryo for transfer and to provide grading decisions (freeze or avoid) for the remaining embryos.
They were then asked to fill in a post-study survey about their experience with the approach.

\subsection{Results}

All four embryologists successfully graded the set of 12 embryos in their sessions, completing assessments from fertilization through day-5 viability decisions.
During the session, participants increasingly relied on automated predictions—such as the number of pronuclei—with rapid visual verification, significantly streamlining their workflows.
Although participants experienced an initial learning curve due to the novelty of the system, completion times for fertilization and timing verifications were comparable to, or shorter than, traditional workflows (between 15 and 40 minutes compared to the typically expected 25 to 40 minutes). This result suggests gains in efficiency.

Participants' interactions with \NAME varied based on individual preferences. P1 extensively used smart playback, frequently toggled visual outlines to inspect morphological features, and reviewed every embryo, though proceeding quickly through embryos that were not visually highlighted. In contrast, the other participants predominantly relied on direct mouse interactions. P3 closely inspected developmental timings, frequently making minor one-frame adjustments. Early identification of non-viable embryos varied: P1 and P2 made confident early decisions, while P3 and P4 delayed their decisions until final comparisons.
All participants mostly used the system in the order of the workflow steps with only occasional backtracking, indicating that the views align with their grading workflow.
P1 mentioned a preference for seeing the embryo scores earlier in the workflow steps.
P2 and P4 interacted extensively with the final comparison view and liked that they could directly compare a subset of embryos in a 'less crowded way', whereas P1 and P3 allocated more attention and time to the early workflow phases.
Occasionally, workflow interruptions reduced efficiency, partly because participants used their own equipment, often with comparatively small laptop screens.

All participants greatly appreciated the color-coded feature visualizations.
The feature contribution charts explaining the classifier score were particularly appreciated, as they clearly illustrated the positive and negative influences on each embryo’s evaluation.
This transparency was especially useful during difficult freeze decisions, allowing participants to understand, discuss, or reconsider the system's scores rather than simply dismissing or accepting them.
Viability scoring generally proved to be instrumental for final embryo selection and participants particularly liked how automatically generated scores were deeply integrated with detailed image analyses—something their current systems lack.
P1, P3, and P4 confidently selected the top-scoring embryo for transfer. P2 favored a different embryo, but briefly reconsidered after noticing that the scoring system had assigned it a slightly lower viability score due to morphological factors.

The small overview visualizations, such as pronuclei detection circles next to the tiles, helped all participants focus their analysis, while the context-dependent representative images supported them in verifying automated assessments.
P1 explicitly praised the interactive visual timeline as "very convenient" and liked the quantification of previously subjective criteria such as fragmentation.
However, all participants indicated they would have benefited from a more extended introductory training period to familiarize themselves further with the system and more quickly grasp uncertainty visualizations such as the display of alternative stage predictions.

\section{Discussion}
\label{sec:discussion}

Compared to existing ML-based systems, our approach introduces a novel human-AI collaboration for embryo selection.
In particular, it delivers transparent and contextualized viability scoring for increased trustworthiness (G4).
It further provides visual summaries and explicit guidance on clinically established grading criteria—such as cell symmetry and fragmentation—that have traditionally been subject to subjective interpretation (G2, G3).
This integration empowers embryologists to efficiently and confidently analyze microscopy data, seamlessly blending manual expertise with computational support (G2, G5).
The system enables customizable levels of in-depth analysis and validation, facilitating flexible adaptation to individual workflows and clinical requirements (G1).

A primary limitation of this study is its focus on a single clinical center and reliance on the EmbryoScope imaging system.
Nevertheless, our dataset includes tens of millions of images, capturing a wide variety of imaging conditions—including variations in quality, contrast, and embryo movement—that are representative of real-world clinical diversity seen across different vendors.
Additionally, the design of the workflow and selection of grading features were informed by both broad clinical expertise and an extensive literature review, supporting the method’s relevance and potential adaptability to laboratory practices elsewhere (G1).
Our viability classifier utilizes engineered, abstracted features following a rigorous data cleaning process (App.~C), which suggests generalizability. The system’s use of background processing and small multiples further contributes to its scalability for typical clinical datasets, although the final comparative view is optimized for a focused, preselected embryo subset (G5).

User feedback from our study indicates that the comprehensive grading features facilitated holistic and confident embryo assessment, even for extensive microscopy datasets. Nonetheless, the introduction of new visual analytics tools necessitated a brief adaptation period, with some clinicians quickly adopting automated hinting features while others preferred to independently verify results via manual assessment. The combination of familiar video playback with novel visualizations eased this transition, though an initial learning curve persisted.

While our evaluation offers promising evidence of the system’s utility and efficiency, broader validation—particularly through multi-center, prospective clinical trials—remains essential. Lastly, it is important to note that the current classifier models correlations rather than causal relationships and assumes independence among embryos, without accounting for intra-cycle dependencies.

\section{Conclusion and Future Work}

In this paper, we introduced \emph{\NAME}, a visual analytics system that integrates advanced machine learning techniques with interactive visualizations, developed collaboratively with embryologists, biologists, and ML researchers, aimed at facilitating embryo selection in IVF treatment. By seamlessly combining automated embryo annotation with interpretable visual analytics, our system addresses critical clinical needs that current manual or opaque computational methods fail to adequately support.
Future enhancements include developing schematic morphological visualizations that smooth variations across multiple consecutive frames, enabling averaged cell plots of specific stages. Moreover, improving cell-tracking techniques, particularly given the challenges posed by overlapping cells and limited imaging resolution, will deliver richer temporal insights into embryo morphology dynamics. Another prospective improvement is enabling quick, automated visual confirmation of critical developmental timings without requiring full video playback, thus further streamlining clinical workflows.
Addressing modeling improvements and visualization extensions will strengthen the clinical relevance, practical utility, and interpretability of future iterations of \emph{\NAME}, facilitating adoption and contributing towards improved IVF treatment outcomes.

\section*{Supplemental Material}
\begin{itemize}
\item The Appendix provides additional details about our iterative design cycles, as well as the methodology used for data processing and machine learning modeling.
\item The supplemental video demonstrates the functionality of our system on a concrete dataset.
\end{itemize}

\acknowledgments{%
The authors would like to thank Yael Kalma for her invaluable support and assistance with this work.
  This work was partially supported by the Harvard Data Science Initiative Postdoctoral Fellowship, NIH grant R01HD104969, and NIH grant 1U01CA284207.%
}

 \bibliographystyle{abbrv-doi-hyperref-narrow-et-al}

\bibliography{references}

\begin{thebibliography}{10}
\renewcommand*{\sfdefault}{PTSansNarrow-TLF}

\bibitem{EmbryoScopeUserManual}
{EmbryoScope} {User} {Manual}.

\bibitem{afnanInterpretableNotBlackbox2021}
M.~A.~M. Afnan, Y.~Liu, V.~Conitzer, et~al.
\newblock Interpretable, not black-box, artificial intelligence should be used
  for embryo selection.
\newblock {\em Human Reproduction Open}, 2021(4):hoab040, Sept. 2021.
  \href{https://doi.org/10.1093/hropen/hoab040}
{doi: \textsf{%
10\hspace{.1pt}\discretionary{.}{%
}{.}\hspace{.4pt}1093\discretionary{/}{%
}{/}hropen\discretionary{/}{%
}{/}hoab040}}


\bibitem{afzal2023visualization}
S.~Afzal, S.~Ghani, M.~M. Hittawe, et~al.
\newblock Visualization and visual analytics approaches for image and video
  datasets: {A} survey.
\newblock {\em ACM Transactions on Interactive Intelligent Systems},
  13(1):1--41, 2023.
\newblock Publisher: ACM New York, NY.

\bibitem{arakiCorrelationPronucleusSize2018}
E.~Araki, F.~Itoi, H.~Honnma, et~al.
\newblock Correlation between the pronucleus size and the potential for human
  single pronucleus zygotes to develop into blastocysts.
\newblock {\em Journal of Assisted Reproduction and Genetics}, 35(5):817--823,
  May 2018. \href{https://doi.org/10.1007/s10815-018-1137-1}
{doi: \textsf{%
10\hspace{.1pt}\discretionary{.}{%
}{.}\hspace{.4pt}1007\discretionary{/}{%
}{/}s10815\discretionary{%
}{-}{-}018\discretionary{%
}{-}{-}1137\discretionary{%
}{-}{-}1}}


\bibitem{arias-hernandezPairAnalyticsCapturing2011}
R.~Arias-Hernandez, L.~T. Kaastra, T.~M. Green, and B.~Fisher.
\newblock Pair {Analytics}: {Capturing} {Reasoning} {Processes} in
  {Collaborative} {Visual} {Analytics}.
\newblock In {\em 2011 44th {Hawaii} {International} {Conference} on {System}
  {Sciences}}, pp. 1--10, Jan. 2011.
\newblock ISSN: 1530-1605. \href{https://doi.org/10.1109/HICSS.2011.339}
{doi: \textsf{%
10\hspace{.1pt}\discretionary{.}{%
}{.}\hspace{.4pt}1109\discretionary{/}{%
}{/}HICSS\hspace{.1pt}\discretionary{.}{%
}{.}\hspace{.4pt}2011\hspace{.1pt}\discretionary{.}{%
}{.}\hspace{.4pt}339}}


\bibitem{baczkowskiMethodsEmbryoScoring2004}
T.~Baczkowski, R.~Kurzawa, and W.~Głabowski.
\newblock Methods of embryo scoring in in vitro fertilization.
\newblock {\em Reproductive biology}, 4(1):5--22, Mar. 2004.
\newblock Place: Poland.

\bibitem{beyer2007high}
J.~Beyer, M.~Hadwiger, S.~Wolfsberger, and K.~Bühler.
\newblock High-quality multimodal volume rendering for preoperative planning of
  neurosurgical interventions.
\newblock {\em IEEE Transactions on Visualization and Computer Graphics},
  13(6):1696--1703, 2007.
\newblock Publisher: IEEE.

\bibitem{bormannPerformanceDeepLearning2020}
C.~L. Bormann, M.~K. Kanakasabapathy, P.~Thirumalaraju, et~al.
\newblock Performance of a deep learning based neural network in the selection
  of human blastocysts for implantation.
\newblock {\em eLife}, 9:e55301, Sept. 2020.
\newblock Publisher: eLife Sciences Publications, Ltd.
  \href{https://doi.org/10.7554/eLife.55301}
{doi: \textsf{%
10\hspace{.1pt}\discretionary{.}{%
}{.}\hspace{.4pt}7554\discretionary{/}{%
}{/}eLife\hspace{.1pt}\discretionary{.}{%
}{.}\hspace{.4pt}55301}}


\bibitem{bormannConsistencyObjectivityAutomated2020}
C.~L. Bormann, P.~Thirumalaraju, M.~K. Kanakasabapathy, et~al.
\newblock Consistency and objectivity of automated embryo assessments using
  deep neural networks.
\newblock {\em Fertility and Sterility}, 113(4):781--787.e1, Apr. 2020.
  \href{https://doi.org/10.1016/j.fertnstert.2019.12.004}
{doi: \textsf{%
10\hspace{.1pt}\discretionary{.}{%
}{.}\hspace{.4pt}1016\discretionary{/}{%
}{/}j\hspace{.1pt}\discretionary{.}{%
}{.}\hspace{.4pt}fertnstert\hspace{.1pt}\discretionary{.}{%
}{.}\hspace{.4pt}2019\hspace{.1pt}\discretionary{.}{%
}{.}\hspace{.4pt}12\hspace{.1pt}\discretionary{.}{%
}{.}\hspace{.4pt}004}}


\bibitem{brumarTypologyDecisionMakingTasks2024}
C.~D. Brumar, S.~Molnar, G.~Appleby, et~al.
\newblock A {Typology} of {Decision}-{Making} {Tasks} for {Visualization}.
\newblock {\em arXiv preprint arXiv:2404.08812}, 2024.

\bibitem{buiTimeLineVisualizingIntegrated2007}
A.~A.~T. Bui, D.~R. Aberle, and H.~Kangarloo.
\newblock {TimeLine}: {Visualizing} {Integrated} {Patient} {Records}.
\newblock {\em IEEE Transactions on Information Technology in Biomedicine},
  11(4):462--473, July 2007.
\newblock Conference Name: IEEE Transactions on Information Technology in
  Biomedicine. \href{https://doi.org/10.1109/TITB.2006.884365}
{doi: \textsf{%
10\hspace{.1pt}\discretionary{.}{%
}{.}\hspace{.4pt}1109\discretionary{/}{%
}{/}TITB\hspace{.1pt}\discretionary{.}{%
}{.}\hspace{.4pt}2006\hspace{.1pt}\discretionary{.}{%
}{.}\hspace{.4pt}884365}}


\bibitem{carpenterCellProfilerImageAnalysis2006}
A.~E. Carpenter, T.~R. Jones, M.~R. Lamprecht, et~al.
\newblock {CellProfiler}: image analysis software for identifying and
  quantifying cell phenotypes.
\newblock {\em Genome Biology}, 7(10):R100, Oct. 2006.
  \href{https://doi.org/10.1186/gb-2006-7-10-r100}
{doi: \textsf{%
10\hspace{.1pt}\discretionary{.}{%
}{.}\hspace{.4pt}1186\discretionary{/}{%
}{/}gb\discretionary{%
}{-}{-}2006\discretionary{%
}{-}{-}7\discretionary{%
}{-}{-}10\discretionary{%
}{-}{-}r100}}


\bibitem{cedilnikIntegrationInformationVolume2008}
A.~Cedilnik, J.~Baumes, L.~Ibanez, et~al.
\newblock Integration of information and volume visualization for analysis of
  cell lineage and gene expression during embryogenesis.
\newblock In {\em Visualization and {Data} {Analysis} 2008}, vol. 6809, pp.
  193--203. SPIE, Jan. 2008. \href{https://doi.org/10.1117/12.768014}
{doi: \textsf{%
10\hspace{.1pt}\discretionary{.}{%
}{.}\hspace{.4pt}1117\discretionary{/}{%
}{/}12\hspace{.1pt}\discretionary{.}{%
}{.}\hspace{.4pt}768014}}


\bibitem{chandra2013infertility}
A.~Chandra, C.~E. Copen, and E.~H. Stephen.
\newblock {\em Infertility and impaired fecundity in the {United} {States},
  1982-2010: data from the {National} {Survey} of {Family} {Growth}}.
\newblock US Department of Health and Human Services, Centers for Disease
  Control and …, 2013.
\newblock Number: 2013.

\bibitem{chandraInfertilityImpairedFecundity2013}
A.~Chandra and E.~H. Stephen.
\newblock Infertility and {Impaired} {Fecundity} in the {United} {States},
  1982–2010: {Data} {From} the {National} {Survey} of {Family} {Growth}.
\newblock (67), 2013.

\bibitem{chavez-badiolaPredictingPregnancyTest2020}
A.~Chavez-Badiola, A.~Flores-Saiffe~Farias, G.~Mendizabal-Ruiz, et~al.
\newblock Predicting pregnancy test results after embryo transfer by image
  feature extraction and analysis using machine learning.
\newblock {\em Scientific Reports}, 10(1):4394, Mar. 2020.
\newblock Publisher: Nature Publishing Group.
  \href{https://doi.org/10.1038/s41598-020-61357-9}
{doi: \textsf{%
10\hspace{.1pt}\discretionary{.}{%
}{.}\hspace{.4pt}1038\discretionary{/}{%
}{/}s41598\discretionary{%
}{-}{-}020\discretionary{%
}{-}{-}61357\discretionary{%
}{-}{-}9}}


\bibitem{chenXGBoostScalableTree2016}
T.~Chen and C.~Guestrin.
\newblock {XGBoost}: {A} {Scalable} {Tree} {Boosting} {System}.
\newblock In {\em Proceedings of the 22nd {ACM} {SIGKDD} {International}
  {Conference} on {Knowledge} {Discovery} and {Data} {Mining}}, {KDD} '16, pp.
  785--794. Association for Computing Machinery, New York, NY, USA, Aug. 2016.
  \href{https://doi.org/10.1145/2939672.2939785}
{doi: \textsf{%
10\hspace{.1pt}\discretionary{.}{%
}{.}\hspace{.4pt}1145\discretionary{/}{%
}{/}2939672\hspace{.1pt}\discretionary{.}{%
}{.}\hspace{.4pt}2939785}}


\bibitem{chengVBridgeConnectingDots2022}
F.~Cheng, D.~Liu, F.~Du, et~al.
\newblock {VBridge}: {Connecting} the {Dots} {Between} {Features} and {Data} to
  {Explain} {Healthcare} {Models}.
\newblock {\em IEEE Transactions on Visualization and Computer Graphics},
  28(1):378--388, Jan. 2022. \href{https://doi.org/10.1109/TVCG.2021.3114836}
{doi: \textsf{%
10\hspace{.1pt}\discretionary{.}{%
}{.}\hspace{.4pt}1109\discretionary{/}{%
}{/}TVCG\hspace{.1pt}\discretionary{.}{%
}{.}\hspace{.4pt}2021\hspace{.1pt}\discretionary{.}{%
}{.}\hspace{.4pt}3114836}}


\bibitem{cruzTimingCellDivision2012}
M.~Cruz, N.~Garrido, J.~Herrero, et~al.
\newblock Timing of cell division in human cleavage-stage embryos is linked
  with blastocyst formation and quality.
\newblock {\em Reproductive BioMedicine Online}, 25(4):371--381, Oct. 2012.
  \href{https://doi.org/10.1016/j.rbmo.2012.06.017}
{doi: \textsf{%
10\hspace{.1pt}\discretionary{.}{%
}{.}\hspace{.4pt}1016\discretionary{/}{%
}{/}j\hspace{.1pt}\discretionary{.}{%
}{.}\hspace{.4pt}rbmo\hspace{.1pt}\discretionary{.}{%
}{.}\hspace{.4pt}2012\hspace{.1pt}\discretionary{.}{%
}{.}\hspace{.4pt}06\hspace{.1pt}\discretionary{.}{%
}{.}\hspace{.4pt}017}}


\bibitem{dinklaScreenitVisualAnalysis2017}
K.~Dinkla, H.~Strobelt, B.~Genest, et~al.
\newblock Screenit: {Visual} {Analysis} of {Cellular} {Screens}.
\newblock {\em IEEE Transactions on Visualization and Computer Graphics},
  23(1):591--600, Jan. 2017. \href{https://doi.org/10.1109/TVCG.2016.2598587}
{doi: \textsf{%
10\hspace{.1pt}\discretionary{.}{%
}{.}\hspace{.4pt}1109\discretionary{/}{%
}{/}TVCG\hspace{.1pt}\discretionary{.}{%
}{.}\hspace{.4pt}2016\hspace{.1pt}\discretionary{.}{%
}{.}\hspace{.4pt}2598587}}


\bibitem{divakaranAugmentingFastforwardRewind2005}
A.~Divakaran, C.~Forlines, T.~Lanning, et~al.
\newblock Augmenting fast-forward and rewind for personal digital video
  recorders.
\newblock In {\em 2005 {Digest} of {Technical} {Papers}. {International}
  {Conference} on {Consumer} {Electronics}, 2005. {ICCE}.}, pp. 43--44, Jan.
  2005.
\newblock ISSN: 2158-4001. \href{https://doi.org/10.1109/ICCE.2005.1429708}
{doi: \textsf{%
10\hspace{.1pt}\discretionary{.}{%
}{.}\hspace{.4pt}1109\discretionary{/}{%
}{/}ICCE\hspace{.1pt}\discretionary{.}{%
}{.}\hspace{.4pt}2005\hspace{.1pt}\discretionary{.}{%
}{.}\hspace{.4pt}1429708}}


\bibitem{duffyGlyphbasedVideoVisualization2015}
B.~Duffy, J.~Thiyagalingam, S.~Walton, et~al.
\newblock Glyph-based video visualization for semen analysis.
\newblock {\em IEEE Transactions on Visualization and Computer Graphics},
  21(8):980--993, 2015. \href{https://doi.org/10.1109/TVCG.2013.265}
{doi: \textsf{%
10\hspace{.1pt}\discretionary{.}{%
}{.}\hspace{.4pt}1109\discretionary{/}{%
}{/}TVCG\hspace{.1pt}\discretionary{.}{%
}{.}\hspace{.4pt}2013\hspace{.1pt}\discretionary{.}{%
}{.}\hspace{.4pt}265}}


\bibitem{elderInVitroFertilization2010}
K.~Elder and B.~Dale.
\newblock {\em In-{Vitro} {Fertilization}}.
\newblock Cambridge University Press, Cambridge, 3 ed., 2010.
  \href{https://doi.org/10.1017/CBO9780511984761}
{doi: \textsf{%
10\hspace{.1pt}\discretionary{.}{%
}{.}\hspace{.4pt}1017\discretionary{/}{%
}{/}CBO9780511984761}}


\bibitem{ezoeCharacteristicsCytoplasmicHalo2021}
K.~Ezoe, T.~Miki, T.~Okimura, et~al.
\newblock Characteristics of the cytoplasmic halo during fertilisation
  correlate with the live birth rate after fresh cleaved embryo transfer on day
  2 in minimal ovarian stimulation cycles: a retrospective observational study.
\newblock {\em Reproductive Biology and Endocrinology}, 19(1):172, Nov. 2021.
  \href{https://doi.org/10.1186/s12958-021-00859-1}
{doi: \textsf{%
10\hspace{.1pt}\discretionary{.}{%
}{.}\hspace{.4pt}1186\discretionary{/}{%
}{/}s12958\discretionary{%
}{-}{-}021\discretionary{%
}{-}{-}00859\discretionary{%
}{-}{-}1}}


\bibitem{fancsovitsEarlyPronuclearBreakdown2005}
P.~Fancsovits, L.~Toth, Z.~F. Takacs, et~al.
\newblock Early pronuclear breakdown is a good indicator of embryo quality and
  viability.
\newblock {\em Fertility and Sterility}, 84(4):881--887, Oct. 2005.
  \href{https://doi.org/10.1016/j.fertnstert.2005.03.068}
{doi: \textsf{%
10\hspace{.1pt}\discretionary{.}{%
}{.}\hspace{.4pt}1016\discretionary{/}{%
}{/}j\hspace{.1pt}\discretionary{.}{%
}{.}\hspace{.4pt}fertnstert\hspace{.1pt}\discretionary{.}{%
}{.}\hspace{.4pt}2005\hspace{.1pt}\discretionary{.}{%
}{.}\hspace{.4pt}03\hspace{.1pt}\discretionary{.}{%
}{.}\hspace{.4pt}068}}


\bibitem{fangerauSimilarityAnalysisCell2012}
J.~Fangerau, B.~Höckendorf, J.~Wittbrodt, and H.~Leitte.
\newblock Similarity analysis of cell movements in video microscopy.
\newblock In {\em 2012 {IEEE} {Symposium} on {Biological} {Data}
  {Visualization} ({BioVis})}, pp. 69--76, Oct. 2012.
  \href{https://doi.org/10.1109/BioVis.2012.6378595}
{doi: \textsf{%
10\hspace{.1pt}\discretionary{.}{%
}{.}\hspace{.4pt}1109\discretionary{/}{%
}{/}BioVis\hspace{.1pt}\discretionary{.}{%
}{.}\hspace{.4pt}2012\hspace{.1pt}\discretionary{.}{%
}{.}\hspace{.4pt}6378595}}


\bibitem{floricelRosesHaveThorns2024}
C.~Floricel, A.~Wentzel, A.~Mohamed, et~al.
\newblock Roses {Have} {Thorns}: {Understanding} the {Downside} of
  {Oncological} {Care} {Delivery} {Through} {Visual} {Analytics} and
  {Sequential} {Rule} {Mining}.
\newblock {\em IEEE Transactions on Visualization and Computer Graphics},
  30(1):1227--1237, Jan. 2024.
\newblock Conference Name: IEEE Transactions on Visualization and Computer
  Graphics. \href{https://doi.org/10.1109/TVCG.2023.3326939}
{doi: \textsf{%
10\hspace{.1pt}\discretionary{.}{%
}{.}\hspace{.4pt}1109\discretionary{/}{%
}{/}TVCG\hspace{.1pt}\discretionary{.}{%
}{.}\hspace{.4pt}2023\hspace{.1pt}\discretionary{.}{%
}{.}\hspace{.4pt}3326939}}


\bibitem{centersfordiseasecontrolandprevention2021AssistedReproductive2023}
C.~for Disease Control {and}~Prevention.
\newblock {\em 2021 {Assisted} {Reproductive} {Technology} {Fertility} {Clinic}
  and {National} {Summary} {Report}}.
\newblock US Dept of Health and Human Services, 2023.

\bibitem{gardnerBlastocystScoreAffects2000}
D.~K. Gardner, M.~Lane, J.~Stevens, et~al.
\newblock Blastocyst score affects implantation and pregnancy outcome: towards
  a single blastocyst transfer.
\newblock {\em Fertility and Sterility}, 73(6):1155--1158, June 2000.
  \href{https://doi.org/10.1016/S0015-0282(00)00518-5}
{doi: \textsf{%
10\hspace{.1pt}\discretionary{.}{%
}{.}\hspace{.4pt}1016\discretionary{/}{%
}{/}S0015\discretionary{%
}{-}{-}0282\discretionary{%
}{(}{(}00\discretionary{)}{%
}{)}00518\discretionary{%
}{-}{-}5}}


\bibitem{gardnerHumanGametesPreimplantation2013}
D.~K. Gardner, D.~Sakkas, E.~Seli, and D.~Wells, eds.
\newblock {\em Human {Gametes} and {Preimplantation} {Embryos}: {Assessment}
  and {Diagnosis}}.
\newblock Springer, New York, NY, 2013.
  \href{https://doi.org/10.1007/978-1-4614-6651-2}
{doi: \textsf{%
10\hspace{.1pt}\discretionary{.}{%
}{.}\hspace{.4pt}1007\discretionary{/}{%
}{/}978\discretionary{%
}{-}{-}1\discretionary{%
}{-}{-}4614\discretionary{%
}{-}{-}6651\discretionary{%
}{-}{-}2}}


\bibitem{gotz2014decisionflow}
D.~Gotz and H.~Stavropoulos.
\newblock Decisionflow: {Visual} analytics for high-dimensional temporal event
  sequence data.
\newblock {\em IEEE Transactions on Visualization and Computer Graphics},
  20(12):1783--1792, 2014.
\newblock Publisher: IEEE.

\bibitem{gotz2011visual}
D.~Gotz, J.~Sun, N.~Cao, and S.~Ebadollahi.
\newblock Visual cluster analysis in support of clinical decision intelligence.
\newblock In {\em {AMIA} annual symposium proceedings}, vol. 2011, p. 481.
  American Medical Informatics Association, 2011.

\bibitem{guptaObesityPredictionEHR2022}
M.~Gupta, T.-L.~T. Phan, H.~T. Bunnell, and R.~Beheshti.
\newblock Obesity {Prediction} with {EHR} {Data}: {A} {Deep} {Learning}
  {Approach} with {Interpretable} {Elements}.
\newblock {\em ACM Transactions on Computing for Healthcare}, 3(3):32:1--32:19,
  Apr. 2022. \href{https://doi.org/10.1145/3506719}
{doi: \textsf{%
10\hspace{.1pt}\discretionary{.}{%
}{.}\hspace{.4pt}1145\discretionary{/}{%
}{/}3506719}}


\bibitem{troidl2023vimo}
{J. Troidl}, {S. Warchol}, {J. Choi}, et~al.
\newblock {ViMO} - {Visual} {Analysis} of {Neuronal} {Connectivity} {Motifs}.
\newblock {\em IEEE Transactions on Visualization and Computer Graphics},
  30(1):748--758, Jan. 2024. \href{https://doi.org/10.1109/TVCG.2023.3327388}
{doi: \textsf{%
10\hspace{.1pt}\discretionary{.}{%
}{.}\hspace{.4pt}1109\discretionary{/}{%
}{/}TVCG\hspace{.1pt}\discretionary{.}{%
}{.}\hspace{.4pt}2023\hspace{.1pt}\discretionary{.}{%
}{.}\hspace{.4pt}3327388}}


\bibitem{jacobsHowMachinelearningRecommendations2021}
M.~Jacobs, M.~F. Pradier, T.~H. McCoy, et~al.
\newblock How machine-learning recommendations influence clinician treatment
  selections: the example of antidepressant selection.
\newblock {\em Translational Psychiatry}, 11(1):108, Feb. 2021.
  \href{https://doi.org/10.1038/s41398-021-01224-x}
{doi: \textsf{%
10\hspace{.1pt}\discretionary{.}{%
}{.}\hspace{.4pt}1038\discretionary{/}{%
}{/}s41398\discretionary{%
}{-}{-}021\discretionary{%
}{-}{-}01224\discretionary{%
}{-}{-}x}}


\bibitem{jangLearningVectorQuantized2023}
W.-D. Jang, D.~Wei, X.~Zhang, et~al.
\newblock Learning {Vector} {Quantized} {Shape} {Code} for {Amodal}
  {Blastomere} {Instance} {Segmentation}.
\newblock In {\em 2023 {IEEE} 20th {International} {Symposium} on {Biomedical}
  {Imaging} ({ISBI})}, pp. 1--5, Apr. 2023.
\newblock ISSN: 1945-8452.
  \href{https://doi.org/10.1109/ISBI53787.2023.10230774}
{doi: \textsf{%
10\hspace{.1pt}\discretionary{.}{%
}{.}\hspace{.4pt}1109\discretionary{/}{%
}{/}ISBI53787\hspace{.1pt}\discretionary{.}{%
}{.}\hspace{.4pt}2023\hspace{.1pt}\discretionary{.}{%
}{.}\hspace{.4pt}10230774}}


\bibitem{jessup2021scope2screen}
J.~Jessup, R.~Krueger, S.~Warchol, et~al.
\newblock Scope2screen: {Focus}+ context techniques for pathology tumor
  assessment in multivariate image data.
\newblock {\em IEEE Transactions on Visualization and Computer Graphics},
  28(1):259--269, 2021.
\newblock Publisher: IEEE.

\bibitem{jin2020carepre}
Z.~Jin, S.~Cui, S.~Guo, et~al.
\newblock Carepre: {An} intelligent clinical decision assistance system.
\newblock {\em ACM Transactions on Computing for Healthcare}, 1(1):1--20, 2020.
\newblock Publisher: ACM New York, NY, USA.

\bibitem{kellerVitessceIntegrativeVisualization2021}
M.~S. Keller, I.~Gold, C.~McCallum, et~al.
\newblock Vitessce: integrative visualization of multimodal and spatially
  resolved single-cell data.
\newblock {\em Nature Methods}, 22(1):63--67, Jan. 2025.
  \href{https://doi.org/10.1038/s41592-024-02436-x}
{doi: \textsf{%
10\hspace{.1pt}\discretionary{.}{%
}{.}\hspace{.4pt}1038\discretionary{/}{%
}{/}s41592\discretionary{%
}{-}{-}024\discretionary{%
}{-}{-}02436\discretionary{%
}{-}{-}x}}


\bibitem{komlodiKeyFramePreview1998}
A.~Komlodi and G.~Marchionini.
\newblock Key frame preview techniques for video browsing.
\newblock In {\em Proceedings of the third {ACM} conference on {Digital}
  libraries}, {DL} '98, pp. 118--125. Association for Computing Machinery, New
  York, NY, USA, May 1998. \href{https://doi.org/10.1145/276675.276688}
{doi: \textsf{%
10\hspace{.1pt}\discretionary{.}{%
}{.}\hspace{.4pt}1145\discretionary{/}{%
}{/}276675\hspace{.1pt}\discretionary{.}{%
}{.}\hspace{.4pt}276688}}


\bibitem{kruegerFacettoCombiningUnsupervised2020}
R.~Krueger, J.~Beyer, W.-D. Jang, et~al.
\newblock Facetto: {Combining} unsupervised and supervised learning for
  hierarchical phenotype analysis in multi-channel image data.
\newblock {\em IEEE Transactions on Visualization and Computer Graphics},
  26(1):227--237, 2020. \href{https://doi.org/10.1109/TVCG.2019.2934547}
{doi: \textsf{%
10\hspace{.1pt}\discretionary{.}{%
}{.}\hspace{.4pt}1109\discretionary{/}{%
}{/}TVCG\hspace{.1pt}\discretionary{.}{%
}{.}\hspace{.4pt}2019\hspace{.1pt}\discretionary{.}{%
}{.}\hspace{.4pt}2934547}}


\bibitem{langeAardvarkCompositeVisualizations2025}
D.~Lange, R.~Judson-Torres, T.~A. Zangle, and A.~Lex.
\newblock Aardvark: {Composite} {Visualizations} of {Trees}, {Time}-{Series},
  and {Images}.
\newblock {\em IEEE Transactions on Visualization and Computer Graphics},
  31(1):1290--1300, Jan. 2025.
\newblock Conference Name: IEEE Transactions on Visualization and Computer
  Graphics. \href{https://doi.org/10.1109/TVCG.2024.3456193}
{doi: \textsf{%
10\hspace{.1pt}\discretionary{.}{%
}{.}\hspace{.4pt}1109\discretionary{/}{%
}{/}TVCG\hspace{.1pt}\discretionary{.}{%
}{.}\hspace{.4pt}2024\hspace{.1pt}\discretionary{.}{%
}{.}\hspace{.4pt}3456193}}


\bibitem{langeLoonUsingExemplars2022}
D.~Lange, E.~Polanco, R.~Judson-Torres, et~al.
\newblock Loon: {Using} {Exemplars} to {Visualize} {Large}-{Scale} {Microscopy}
  {Data}.
\newblock {\em IEEE Transactions on Visualization and Computer Graphics},
  28(1):248--258, Jan. 2022.
\newblock Conference Name: IEEE Transactions on Visualization and Computer
  Graphics. \href{https://doi.org/10.1109/TVCG.2021.3114766}
{doi: \textsf{%
10\hspace{.1pt}\discretionary{.}{%
}{.}\hspace{.4pt}1109\discretionary{/}{%
}{/}TVCG\hspace{.1pt}\discretionary{.}{%
}{.}\hspace{.4pt}2021\hspace{.1pt}\discretionary{.}{%
}{.}\hspace{.4pt}3114766}}


\bibitem{leahyAutomatedMeasurementsKey2020}
B.~D. Leahy, W.-D. Jang, H.~Y. Yang, et~al.
\newblock Automated {Measurements} of {Key} {Morphological} {Features} of
  {Human} {Embryos} for {IVF}.
\newblock In A.~L. Martel, P.~Abolmaesumi, D.~Stoyanov, et~al., eds., {\em
  Medical {Image} {Computing} and {Computer} {Assisted} {Intervention} –
  {MICCAI} 2020}, pp. 25--35. Springer International Publishing, Cham, 2020.

\bibitem{ledesmaHealthTimelineInsightbased2019}
A.~Ledesma, N.~Bidargaddi, J.~Strobel, et~al.
\newblock Health timeline: an insight-based study of a timeline visualization
  of clinical data.
\newblock {\em BMC Medical Informatics and Decision Making}, 19(1):170, Aug.
  2019. \href{https://doi.org/10.1186/s12911-019-0885-x}
{doi: \textsf{%
10\hspace{.1pt}\discretionary{.}{%
}{.}\hspace{.4pt}1186\discretionary{/}{%
}{/}s12911\discretionary{%
}{-}{-}019\discretionary{%
}{-}{-}0885\discretionary{%
}{-}{-}x}}


\bibitem{lukyanenkoDevelopmentalStageClassification2021}
S.~Lukyanenko, W.-D. Jang, D.~Wei, et~al.
\newblock Developmental {Stage} {Classification} of {Embryos} {Using}
  {Two}-{Stream} {Neural} {Network} with {Linear}-{Chain} {Conditional}
  {Random} {Field}.
\newblock In M.~de~Bruijne, P.~C. Cattin, S.~Cotin, et~al., eds., {\em Medical
  {Image} {Computing} and {Computer} {Assisted} {Intervention} – {MICCAI}
  2021}, pp. 363--372. Springer International Publishing, Cham, 2021.

\bibitem{lundbergLocalExplanationsGlobal2020}
S.~M. Lundberg, G.~Erion, H.~Chen, et~al.
\newblock From local explanations to global understanding with explainable {AI}
  for trees.
\newblock {\em Nature Machine Intelligence}, 2(1):56--67, Jan. 2020.
\newblock Publisher: Nature Publishing Group.
  \href{https://doi.org/10.1038/s42256-019-0138-9}
{doi: \textsf{%
10\hspace{.1pt}\discretionary{.}{%
}{.}\hspace{.4pt}1038\discretionary{/}{%
}{/}s42256\discretionary{%
}{-}{-}019\discretionary{%
}{-}{-}0138\discretionary{%
}{-}{-}9}}


\bibitem{maKeyframeExtractionLaparoscopic2021}
M.~Ma, S.~Mei, S.~Wan, et~al.
\newblock Keyframe {Extraction} {From} {Laparoscopic} {Videos} via {Diverse}
  and {Weighted} {Dictionary} {Selection}.
\newblock {\em IEEE Journal of Biomedical and Health Informatics},
  25(5):1686--1698, May 2021.
\newblock Conference Name: IEEE Journal of Biomedical and Health Informatics.
  \href{https://doi.org/10.1109/JBHI.2020.3019198}
{doi: \textsf{%
10\hspace{.1pt}\discretionary{.}{%
}{.}\hspace{.4pt}1109\discretionary{/}{%
}{/}JBHI\hspace{.1pt}\discretionary{.}{%
}{.}\hspace{.4pt}2020\hspace{.1pt}\discretionary{.}{%
}{.}\hspace{.4pt}3019198}}


\bibitem{mundurKeyframebasedVideoSummarization2006}
P.~Mundur, Y.~Rao, and Y.~Yesha.
\newblock Keyframe-based video summarization using {Delaunay} clustering.
\newblock {\em International Journal on Digital Libraries}, 6(2):219--232, Apr.
  2006. \href{https://doi.org/10.1007/s00799-005-0129-9}
{doi: \textsf{%
10\hspace{.1pt}\discretionary{.}{%
}{.}\hspace{.4pt}1007\discretionary{/}{%
}{/}s00799\discretionary{%
}{-}{-}005\discretionary{%
}{-}{-}0129\discretionary{%
}{-}{-}9}}


\bibitem{nasiriOverviewAvailableMethods2015}
N.~Nasiri and P.~Eftekhari-Yazdi.
\newblock An overview of the available methods for morphological scoring of
  pre-implantation embryos in in vitro fertilization.
\newblock {\em Cell journal}, 16(4):392--405, 2015.
\newblock Place: Iran. \href{https://doi.org/10.22074/cellj.2015.486}
{doi: \textsf{%
10\hspace{.1pt}\discretionary{.}{%
}{.}\hspace{.4pt}22074\discretionary{/}{%
}{/}cellj\hspace{.1pt}\discretionary{.}{%
}{.}\hspace{.4pt}2015\hspace{.1pt}\discretionary{.}{%
}{.}\hspace{.4pt}486}}


\bibitem{nie2012compact}
Y.~Nie, C.~Xiao, H.~Sun, and P.~Li.
\newblock Compact video synopsis via global spatiotemporal optimization.
\newblock {\em IEEE Transactions on Visualization and Computer Graphics},
  19(10):1664--1676, 2012.
\newblock Publisher: IEEE.

\bibitem{pretoriusSurveyVisualizationLive2017}
A.~J. Pretorius, I.~A. Khan, and R.~J. Errington.
\newblock A {Survey} of {Visualization} for {Live} {Cell} {Imaging}.
\newblock {\em Computer Graphics Forum}, 36(1):46--63, 2017.
  \href{https://doi.org/10.1111/cgf.12784}
{doi: \textsf{%
10\hspace{.1pt}\discretionary{.}{%
}{.}\hspace{.4pt}1111\discretionary{/}{%
}{/}cgf\hspace{.1pt}\discretionary{.}{%
}{.}\hspace{.4pt}12784}}


\bibitem{reichmanIncidenceDevelopmentZygotes2010}
D.~E. Reichman, K.~V. Jackson, and C.~Racowsky.
\newblock Incidence and development of zygotes exhibiting abnormal pronuclear
  disposition after identification of two pronuclei at the fertilization check.
\newblock {\em Fertility and Sterility}, 94(3):965--970, Aug. 2010.
\newblock Publisher: Elsevier.
  \href{https://doi.org/10.1016/j.fertnstert.2009.04.018}
{doi: \textsf{%
10\hspace{.1pt}\discretionary{.}{%
}{.}\hspace{.4pt}1016\discretionary{/}{%
}{/}j\hspace{.1pt}\discretionary{.}{%
}{.}\hspace{.4pt}fertnstert\hspace{.1pt}\discretionary{.}{%
}{.}\hspace{.4pt}2009\hspace{.1pt}\discretionary{.}{%
}{.}\hspace{.4pt}04\hspace{.1pt}\discretionary{.}{%
}{.}\hspace{.4pt}018}}


\bibitem{informatics8010012}
N.~Rostamzadeh, S.~S. Abdullah, and K.~Sedig.
\newblock Visual analytics for electronic health records: {A} review.
\newblock {\em Informatics}, 8(1), 2021.
\newblock Number: 12. \href{https://doi.org/10.3390/informatics8010012}
{doi: \textsf{%
10\hspace{.1pt}\discretionary{.}{%
}{.}\hspace{.4pt}3390\discretionary{/}{%
}{/}informatics8010012}}


\bibitem{rubioClinicalValidationEmbryo2014}
I.~Rubio, A.~Galán, Z.~Larreategui, et~al.
\newblock Clinical validation of embryo culture and selection by morphokinetic
  analysis: a randomized, controlled trial of the {EmbryoScope}.
\newblock {\em Fertility and Sterility}, 102(5):1287--1294.e5, Nov. 2014.
  \href{https://doi.org/10.1016/j.fertnstert.2014.07.738}
{doi: \textsf{%
10\hspace{.1pt}\discretionary{.}{%
}{.}\hspace{.4pt}1016\discretionary{/}{%
}{/}j\hspace{.1pt}\discretionary{.}{%
}{.}\hspace{.4pt}fertnstert\hspace{.1pt}\discretionary{.}{%
}{.}\hspace{.4pt}2014\hspace{.1pt}\discretionary{.}{%
}{.}\hspace{.4pt}07\hspace{.1pt}\discretionary{.}{%
}{.}\hspace{.4pt}738}}


\bibitem{schlachter2019state}
M.~Schlachter, R.~G. Raidou, L.~P. Muren, et~al.
\newblock State-of-the-art report: {Visual} computing in radiation therapy
  planning.
\newblock In {\em Computer {Graphics} {Forum}}, vol.~38, pp. 753--779. Wiley
  Online Library, 2019.
\newblock Number: 3.

\bibitem{silver2020datadriven}
D.~H. Silver, M.~Feder, Y.~Gold-Zamir, et~al.
\newblock Data-driven prediction of embryo implantation probability using {IVF}
  time-lapse imaging.
\newblock In {\em Medical imaging with deep learning}, 2020.

\bibitem{theilgaardlassenDevelopmentValidationDeep2023}
J.~Theilgaard~Lassen, M.~Fly~Kragh, J.~Rimestad, et~al.
\newblock Development and validation of deep learning based embryo selection
  across multiple days of transfer.
\newblock {\em Scientific Reports}, 13(1):4235, Mar. 2023.
\newblock Publisher: Nature Publishing Group.
  \href{https://doi.org/10.1038/s41598-023-31136-3}
{doi: \textsf{%
10\hspace{.1pt}\discretionary{.}{%
}{.}\hspace{.4pt}1038\discretionary{/}{%
}{/}s41598\discretionary{%
}{-}{-}023\discretionary{%
}{-}{-}31136\discretionary{%
}{-}{-}3}}


\bibitem{tranDeepLearningPredictive2019}
D.~Tran, S.~Cooke, P.~J. Illingworth, and D.~K. Gardner.
\newblock Deep learning as a predictive tool for fetal heart pregnancy
  following time-lapse incubation and blastocyst transfer.
\newblock {\em Human Reproduction}, 34(6):1011--1018, May 2019.
  \href{https://doi.org/10.1093/humrep/dez064}
{doi: \textsf{%
10\hspace{.1pt}\discretionary{.}{%
}{.}\hspace{.4pt}1093\discretionary{/}{%
}{/}humrep\discretionary{/}{%
}{/}dez064}}


\bibitem{verbergWhyCouplesDropout2008}
M.~Verberg, M.~Eijkemans, E.~Heijnen, et~al.
\newblock Why do couples drop-out from {IVF} treatment? {A} prospective cohort
  study.
\newblock {\em Human Reproduction}, 23(9):2050--2055, Sept. 2008.
  \href{https://doi.org/10.1093/humrep/den219}
{doi: \textsf{%
10\hspace{.1pt}\discretionary{.}{%
}{.}\hspace{.4pt}1093\discretionary{/}{%
}{/}humrep\discretionary{/}{%
}{/}den219}}


\bibitem{vermilyeaDevelopmentArtificialIntelligencebased2020}
M.~VerMilyea, J.~M.~M. Hall, S.~M. Diakiw, et~al.
\newblock Development of an artificial intelligence-based assessment model for
  prediction of embryo viability using static images captured by optical light
  microscopy during {IVF}.
\newblock {\em Human Reproduction}, 35(4):770--784, Apr. 2020.
  \href{https://doi.org/10.1093/humrep/deaa013}
{doi: \textsf{%
10\hspace{.1pt}\discretionary{.}{%
}{.}\hspace{.4pt}1093\discretionary{/}{%
}{/}humrep\discretionary{/}{%
}{/}deaa013}}


\bibitem{warcholVisinityVisualSpatial2022}
S.~Warchol, R.~Krueger, A.~J. Nirmal, et~al.
\newblock Visinity: {Visual} {Spatial} {Neighborhood} {Analysis} for
  {Multiplexed} {Tissue} {Imaging} {Data}.
\newblock {\em IEEE Transactions on Visualization and Computer Graphics}, pp.
  1--11, 2022. \href{https://doi.org/10.1109/TVCG.2022.3209378}
{doi: \textsf{%
10\hspace{.1pt}\discretionary{.}{%
}{.}\hspace{.4pt}1109\discretionary{/}{%
}{/}TVCG\hspace{.1pt}\discretionary{.}{%
}{.}\hspace{.4pt}2022\hspace{.1pt}\discretionary{.}{%
}{.}\hspace{.4pt}3209378}}


\bibitem{wentzelExplainableSpatialClustering2020}
A.~Wentzel, G.~Canahuate, L.~V. van Dijk, et~al.
\newblock Explainable {Spatial} {Clustering}: {Leveraging} {Spatial} {Data} in
  {Radiation} {Oncology}.
\newblock In {\em 2020 {IEEE} {Visualization} {Conference} ({VIS})}, pp.
  281--285, Oct. 2020. \href{https://doi.org/10.1109/VIS47514.2020.00063}
{doi: \textsf{%
10\hspace{.1pt}\discretionary{.}{%
}{.}\hspace{.4pt}1109\discretionary{/}{%
}{/}VIS47514\hspace{.1pt}\discretionary{.}{%
}{.}\hspace{.4pt}2020\hspace{.1pt}\discretionary{.}{%
}{.}\hspace{.4pt}00063}}


\bibitem{yangBlastAssistDeepLearning2024}
H.~Y. Yang, B.~D. Leahy, W.-D. Jang, et~al.
\newblock {BlastAssist}: a deep learning pipeline to measure interpretable
  features of human embryos.
\newblock {\em Human Reproduction}, p. deae024, Feb. 2024.
  \href{https://doi.org/10.1093/humrep/deae024}
{doi: \textsf{%
10\hspace{.1pt}\discretionary{.}{%
}{.}\hspace{.4pt}1093\discretionary{/}{%
}{/}humrep\discretionary{/}{%
}{/}deae024}}


\bibitem{zajacClinicianFacingAIWild2023}
H.~D. Zając, D.~Li, X.~Dai, et~al.
\newblock Clinician-{Facing} {AI} in the {Wild}: {Taking} {Stock} of the
  {Sociotechnical} {Challenges} and {Opportunities} for {HCI}.
\newblock {\em ACM Transactions on Computer-Human Interaction},
  30(2):33:1--33:39, Mar. 2023. \href{https://doi.org/10.1145/3582430}
{doi: \textsf{%
10\hspace{.1pt}\discretionary{.}{%
}{.}\hspace{.4pt}1145\discretionary{/}{%
}{/}3582430}}


\bibitem{zaninovicArtificialIntelligenceHuman2020}
N.~Zaninovic and Z.~Rosenwaks.
\newblock Artificial intelligence in human in vitro fertilization and
  embryology.
\newblock {\em Fertility and Sterility}, 114(5):914--920, 2020.
  \href{https://doi.org/10.1016/j.fertnstert.2020.09.157}
{doi: \textsf{%
10\hspace{.1pt}\discretionary{.}{%
}{.}\hspace{.4pt}1016\discretionary{/}{%
}{/}j\hspace{.1pt}\discretionary{.}{%
}{.}\hspace{.4pt}fertnstert\hspace{.1pt}\discretionary{.}{%
}{.}\hspace{.4pt}2020\hspace{.1pt}\discretionary{.}{%
}{.}\hspace{.4pt}09\hspace{.1pt}\discretionary{.}{%
}{.}\hspace{.4pt}157}}


\bibitem{zhang2018idmvis}
Y.~Zhang, K.~Chanana, and C.~Dunne.
\newblock {IDMVis}: {Temporal} event sequence visualization for type 1 diabetes
  treatment decision support.
\newblock {\em IEEE Transactions on Visualization and Computer Graphics},
  25(1):512--522, 2018.
\newblock Publisher: IEEE.

\end{thebibliography}

\appendix
\section{Design Iterations}
\label{app:designIterations}

\begin{figure*}
\centering
\includegraphics[width=1.0\linewidth]{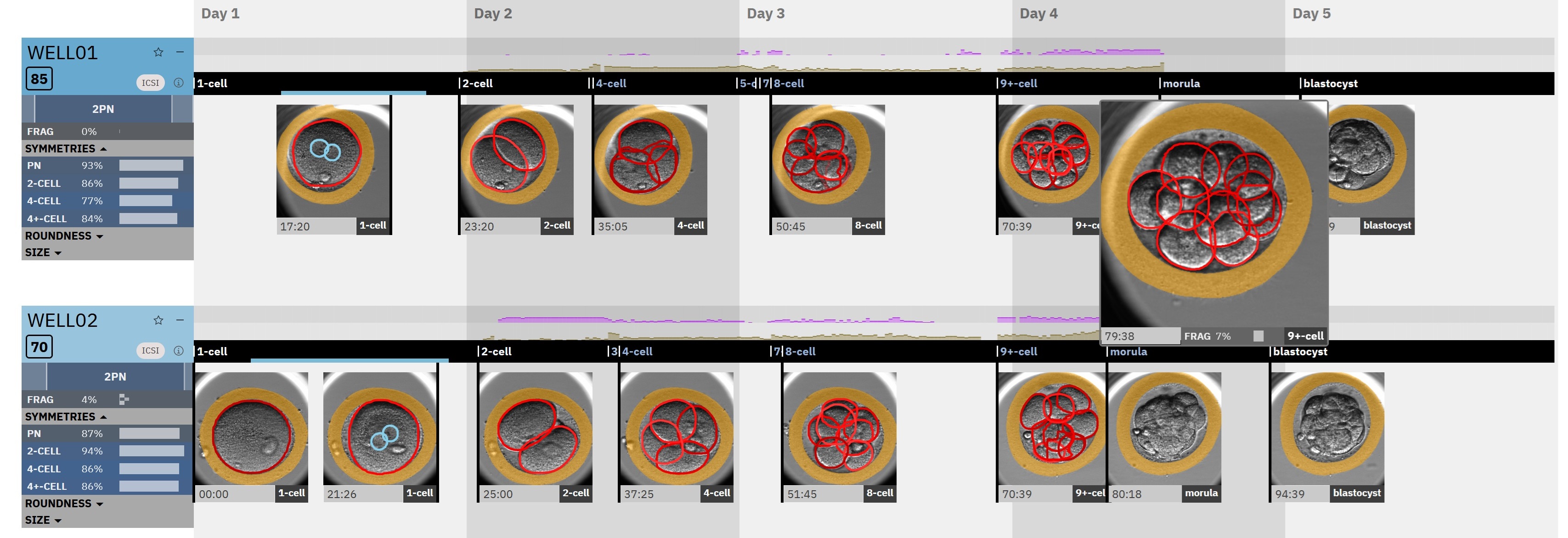}
\caption{\textbf{Initial Design of EmbryoProfiler.} The first version of EmbryoProfiler, developed after collaborative sessions with clinical experts, prioritized comparative timing analyses and supported the selection of embryos for implantation through a ranking-driven interface.}
\label{fig:timeline2024}
\end{figure*}

We conducted iterative design cycles through regular online meetings that involved computational biologists from Harvard University and clinical experts from the Tel Aviv Souraski Medical Center. Clinical collaborators included senior embryologists and the laboratory's principal investigator. These meetings significantly shaped the evolution of \NAME, ensuring our approach matched real-world clinical requirements.

Our initial design, developed after preliminary collaborative sessions, emphasized comparative timing analyses due to their centrality in embryo grading (Fig.~\ref{fig:timeline2024}).
The design specifically targeted support for clinicians’ primary task: identifying and selecting the most promising embryos for implantation.
Additionally, our collaborators highlighted the value of incorporating data-driven modeling into the system to promote consistency and objectivity in embryo assessments.

Thus, our first prototype for \NAME focused heavily on visually ranking embryos side-by-side based on viability predictions, facilitating streamlined yet informed clinical transfer decisions. To explicitly communicate correlations among the identified morphological features, developmental stages, and viability scores, we employed distinct color-coding schemes and overlay visualizations of segmented cell boundaries on embryo image frames.

While clinicians appreciated the enhanced visualization and clear presentation of the recorded videos and the corresponding embryo viability scores, preliminary usability testing revealed several areas for improvement. Users faced challenges when adjusting timing parameters through context-menus, interpreting numeric feature scores such as 80\% symmetry, and dealing with considerable visual density, as the initial design was more suitable for data analysts and biologists than clinical users.

Based on these insights, the following meetings centered around restructuring \NAME into clearly defined sections corresponding directly to key clinical workflow steps, such as fertilization assessment.
After several iterations, this lead to the final revised version presented in this paper.
This included developing more specialized visualizations tailored explicitly for each analytical task to mitigate cognitive overload.
For instance, we developed cell shape plots to better support visual assessments of cell morphology without frame overlays that occlude parts of the image.
We further combined traditional frame-by-frame embryo viewers with summary visualizations that facilitate faster and more intuitive analyses.
To address clinicians' preference for qualitative rather than numeric data, we shifted from displaying numeric scores toward qualitative indicators with our slide-style plots, categorizing embryo quality assessments as good, neutral, or poor. Furthermore, we standardized and explicitly linked color schemes with specific developmental stages, significantly improving the visual interpretability of developmental progress and providing immediate insights.

\section{Preprocessing and Feature Engineering}
\label{app:preprocessing}

We first employ computer vision models to annotate individual frames with raw features, from which we then compute aggregated, grading-relevant, and human-interpretable features for both our visualization interface and the subsequent viability classifier.

\subsection{Frame-By-Frame Predictions}

For each timestep, comprising several focal planes, we obtain the following predictions:

\para{Pixelwise segmentation} of all focal plane images into outside well, inside well (petri dish), zona pellucida (protective shell of ovum), and ovum (egg).

\para{Bounding box} of the actual embryo within the well for each focal plane based on the zona segmentation.

\para{Probability that the embryo is in a certain stage} for each of the twelve stages (empty, 1-cell, ..., 9+-cell, morula, blastocyst, plus unknown).

\para{Blastocyst segmentation} for frames belonging to the blastocyst stage. We obtain segmented areas for inner cell mass, trophectoderm, and blastocoel cavity of the main focal plane F0.

\para{Level of blastomere fragmentation}, a number that typically varies between 0 and 3, with values over 2.5 corresponding to the clinical definition of having a level of fragmentation that exceeds 25\%.

\para{Detected pronuclei and blastomeres}, comprising the path of the polygon and a confidence value.

Following these raw predictions, a level of data cleaning is necessary before feature extraction and automated scoring.

\subsection{Data Cleaning}

In some cases, these CV models may give inaccurate predictions, particularly for highly fragmented embryos.
These incorrect predictions usually fall into three categories: duplicates, other phenomenons mistaken for blastomeres (especially outside the ovum), and noise (typically resulting in weirdly shaped polygons).
We, therefore, discard blastomere and pronucleus predictions that are clearly outside of the ovum based on the computed segmentation.
We also eliminate polygons that are clearly duplicates, having nearly the same area and centroid, and those that do not exhibit an ellipsoid-like shape (area significantly smaller than what the circumference would imply for a circle-like shape).
Finally, we discard low-confidence predictions and pronuclei instances that do not appear in the previous or next frame.

\subsection{Feature Engineering}

After data cleaning, we compute morphological features that clinicians can use to evaluate the viability of embryos for implantation in the corresponding cycle.
We first compute smoothed bounding boxes across focal planes and timesteps so that our interface can zoom onto the actual embryo.
We further compute the following set of features.
They also serve as inputs for our viability classifier (see Sec.~\ref{sec:classifier}), in addition to patient characteristics such as maternal age during ovum pickup and during treatment as well as type of infertility.

\para{Fragmentation:} The median and interquartile range of the level of blastomere fragmentation across all frames in early cleavage stages (2-cell to 9+-cell).

\para{Pronuclei Timings:} We determine when the pronucleus or pronuclei (PN) appear and when they dissolve shortly before the first division.

\para{Stage Assignments and Timings:}
We derive the final monotonically increasing stage assignments by incrementing the stage if the currently estimated stage is higher than before and lower or equal to the predictions of the next three frames.
Using the predicted stage progression, we determine stage timings in reference to the PN fading time to normalize between IVF and ICSI.

\para{Fertilization:} We calculate the likelihoods that the embryo has 0, 1, 2, or 3+ pronuclei by looking at the distribution of the expected number of pronuclei across all frames until the 2-cell stage. We also calculate the distance between pronuclei (if 2+ exist).

\para{Symmetries:} For a given frame, we calculate the per-cell (per-pronucleus) symmetry by averaging the area ratios (<1) with every other cell (pronucleus). The average of this constitutes the symmetry of the current frame. We then derive average pronuclei, 2-cell, 4-cell, and 4+-cell symmetries.

\para{Roundness (Shape):}
For the 2-cell and 4-cell stages, we compute the ratio of the area of the polygon with the expected area of a circle given the observed circumference. This is the isoperimetric quotient that ranges from just above 0 to 1 for a perfect circle.

\para{Zona Width (Thickness):} Using the zona pellucida and ovum segmentations, we derive the median thickness of the zona pellucida and the relation of the thickness to the ovum diameter until the 2-cell stage. We also compute the average thinning of the zona pellucida over time during blastocyst expansion (if present).

\para{Size:} We compute the average area of the detected pronuclei. Using the segmentations, we further determine the areas of the single-cell, ovum, zona pellucida, inner cell mass, and trophectoderm.

\para{Blastocyst Expansion:} Using the segmentations, we use both the ovum expansion and blastocoel expansion per frame in the blastocyst stage to compute a more robust measure of the average blastocoel growth and the peak area of the blastocoel cavity.

\section{Viability Classifier}
\label{app:classifier}

Our embryo viability classifier serves two main purposes.
First, it provides an automatically derived, data-driven score based on human-interpretable features—such as cell-development timings and morphological attributes—that strongly correlate with embryo viability and live birth outcomes. Optionally, patient characteristics, including maternal age and infertility diagnosis, can be incorporated into the classifier. Although consistent across embryos within a single treatment cycle, these patient-level attributes are essential for robust model training due to their confounding relationship with clinical outcomes.

Secondly, the viability classifier enables the visualization and interpretation of the relative influence of individual embryo features on predicted clinical outcomes. Such visualizations facilitate clinical reasoning and guide embryologists to focus attention on the most influential developmental markers. It is important to emphasize that resulting correlations represent observational associations and should not be interpreted causally.

We deliberately chose a machine learning method that utilizes interpretable embryo-derived features. This design promotes transparency and cultivates clinical trust in our scoring and ranking system. Nevertheless, our clinical decision-support approach, centered around visualization, remains broadly model-agnostic and can easily integrate scoring from existing embryo grading methods introduced by clinical facilities.

\subsection{Dataset}
Our dataset consists of $47,336$ embryos captured through tens of millions of EmbryoScope time-lapse microscopy images from the Sourasky Medical Center in Tel Aviv, supplemented by relevant electronic health records (EHR), spanning treatment cycles conducted between 2012 and 2019.
Each embryo is associated with a clinical selection decision: transferred, frozen, avoided, or undecided. We know clinical outcomes for each cycle, specifically live birth occurrences. However, if more embryos got transferred than children born, we do not know which of the transferred embryos was successful.

The length of the microscopy recordings varies significantly within each treatment cycle; embryos selected for early transfer have correspondingly shorter monitoring durations, while non-selected embryos continue incubation.
This introduces potential bias, as incomplete recordings indirectly reflect clinical decisions and thus must not influence derived model inputs. Therefore, careful handling of derived features related to developmental stages was essential to prevent inadvertent information leakage.

EHR patient characteristics employed include maternal age at ovum retrieval, maternal age at time of treatment, measured estradiol hormone levels, and infertility type (primary, secondary, or none). For model training and evaluation, we partition embryos into training, validation, and test subsets following a 60-20-20 patient-level split, ensuring embryos from a single patient do not overlap among these subsets to prevent data leakage. We use larger proportions of validation and test sets since successful outcomes are comparatively rare (only a few embryos are being transferred, and only a few of them lead to successful deliveries).

\subsection{Method}
Our primary objective is to estimate an embryo's likelihood of resulting in a live birth upon transfer. Input features provided to our predictive model consist of interpretable embryo characteristics computed through our computer vision pipeline, supplemented with patient information. In cases where more embryos were transferred than live births obtained, we applied instance-based weighting proportionally to address this imbalance.

A major challenge arises from embryos that are never transferred, which constitute the majority of embryos in our dataset and thus lack known clinical outcomes. Assigning negative outcomes indiscriminately (particularly to frozen embryos) is inappropriate. Conversely, excluding non-transferred embryos from training would significantly narrow our dataset and introduce severe selection bias—excluding many clearly non-viable embryos (e.g., highly fragmented embryos). Another complexity is that many transfers occurred on days 2 or 3 rather than day 5, creating shortened recordings even for embryos clinically deemed viable. To address these challenges, we developed a two-step training approach.

\subsubsection*{Step 1: Maturity Classifier}

We initially developed an embryo maturity classifier to predict embryo developmental potential using expansion status at day 5 as the primary criterion. Embryos were categorized into three distinct maturity stages:

\begin{itemize}
    \item \textbf{Advanced maturity:} Expanded blastocysts
    \item \textbf{Moderate maturity:} Morulae or early-stage blastocysts
    \item \textbf{Low maturity:} Embryos exhibiting minimal or no further developmental progression at day 5
\end{itemize}

These maturity categories closely correlate with declining clinical success rates and can be assessed objectively, significantly reducing the risk of assessor bias.

Our dataset showed that approximately 8\% of embryos classified with advanced maturity were selected for transfer, achieving a live birth success rate of 22\%. Embryos classified as moderate maturity were transferred less frequently (approximately 3\% of embryos), resulting in a slightly lower success rate of about 20\%. In contrast, embryos identified as low maturity were rarely transferred, and no successful births occurred from this category.

The primary objective of creating this classifier was to generate reliable \textit{weak labels} for non-transferred embryos, effectively increasing the usable dataset size for training our primary embryo viability predictor.

For training and evaluating the maturity classifier, we utilized a subset containing $21,709$ embryos continuously monitored until day 5. We deliberately excluded developmental features explicitly related to the maturity-label definition criteria (e.g., blastocyst expansion timings) to prevent label leakage. Class imbalance was addressed by assigning appropriate sample weights prior to model training. The final maturity classifier demonstrated strong predictive performance, attaining an area under the ROC curve (AUC) of $0.89$ on the held-out test dataset.

Importantly, beyond correct categorical assignment, the classifier's probabilistic outputs more accurately reflected underlying embryo viability.
Embryos predicted to have advanced maturity with likelihood scores greater than $0.7$ exhibited enhanced clinical outcomes: 16\% of these embryos were selected for transfer, and approximately one-third resulted in successful births.
This performance surpasses baseline rates observed within comparable embryos categorized by the original ground-truth maturity labels.

\subsubsection*{Step 2: Viability Classifier}
Following maturity classification, we trained our primary viability classifier designed to predict live birth outcomes following embryo transfer. To avoid unintended leakage of clinical decisions into model features, embryo recordings within each treatment cycle were uniformly trimmed to an identical monitoring duration before feature extraction.

We then leveraged the maturity classifier results to enrich our original \emph{training set}. Specifically, frozen or unlabeled embryos strongly classified as demonstrating low maturity (no development or limited development) were added with negative outcomes. Conversely, embryos robustly classified as advanced maturity (expanded blastocysts) were labeled positive, but assigned significantly reduced instance weights (0.05), reflecting considerable uncertainty around their ultimate clinical outcome.
Optimal maturity classifier probability thresholds and hyperparameters were determined using the validation dataset.

Additionally, we balanced positive and negative class weights equally during training to prevent bias toward negative predictions. To guarantee rigorous evaluation, our validation and test subsets comprised exclusively embryos explicitly transferred or clearly labeled as "avoid." With this strategy, we effectively utilized $30,328$ embryos.

We employed an XGBoost model~\cite{chenXGBoostScalableTree2016}, known to deliver state-of-the-art predictive performance for classification tasks involving such tabular data.

\subsection{Results}

\begin{figure}
\centering
\includegraphics[width=1.0\linewidth]{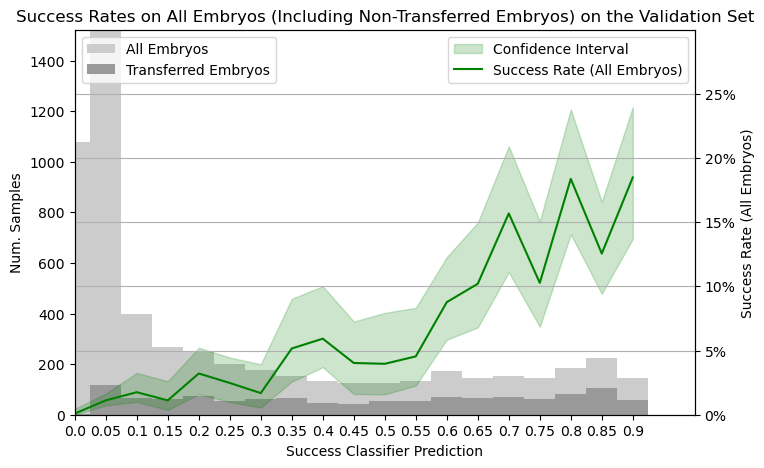}
\caption{\textbf{Success Rates on the Validation Set} with bootstrapped confidence intervals (90\%), computed from successful outcomes among all embryos (including non-transferred embryos) within each score range. The stacked bar chart shows the number of transferred embryos (dark bars) relative to the total number of embryos (light bars).}
\label{fig:successRatesValidation}
\vspace{-0.3cm}
\end{figure}

\begin{figure}
\centering
\includegraphics[width=1.0\linewidth]{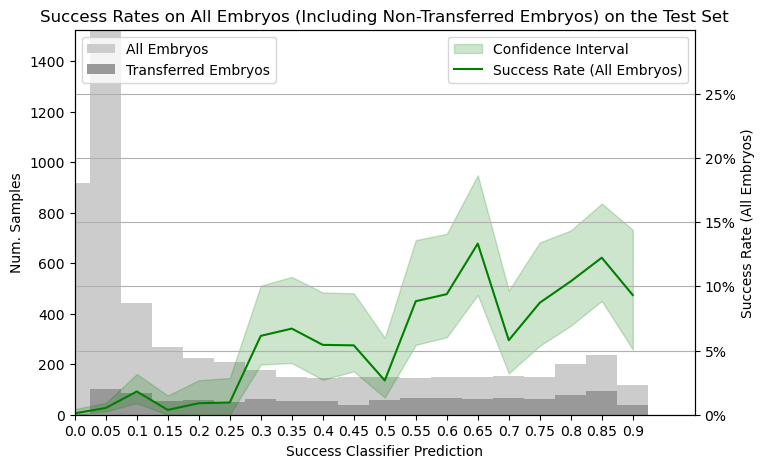}
\caption{\textbf{Success Rates on the Test Set} with bootstrapped confidence intervals (90\%), computed from successful outcomes among all embryos (including non-transferred embryos) within each score range. The stacked bar chart shows the number of transferred embryos (dark bars) relative to the total number of embryos (light bars).}
\label{fig:successRatesTest}
\vspace{-0.3cm}
\end{figure}

Evaluations conducted on our independent test set demonstrate strong viability classifier performance overall, yielding an AUC of $0.83$. Even within the clinical subset representing only embryos pre-screened for transfer (a notably more challenging evaluation setting), our method remains predictive and competitive, attaining an AUC of $0.70$. This highlights the effectiveness of our approach, particularly given the considerable clinical judgment already involved in prior embryo selections.

Fig.\ref{fig:successRatesValidation} and Fig.\ref{fig:successRatesTest} provide more detailed results for the validation and test sets, respectively. For each predicted likelihood score interval, we compute the proportion of successful clinical outcomes (deliveries) across the entire set of embryos scored in that interval, including those embryos not selected for transfer (which are therefore counted as unsuccessful outcomes). Confidence intervals are computed using a bootstrap procedure at a confidence level of 0.9.

The results indicate that higher predicted scores correlate positively with increased embryo viability, reflected both in higher transfer proportions and overall success rates. Although the absolute number of transfers remains comparatively stable across score intervals, this does not imply a frequent disagreement between classifier predictions and historical clinical decisions. Rather, it likely reflects the substantial variability in average embryo viability among cycles, where clinicians frequently face limited availability of highly viable embryos. In these scenarios, our data-driven scores could potentially assist clinicians in determining whether initiating new cycles is beneficial or necessary.

Overall, our proposed two-stage training framework demonstrates robust predictive accuracy for clinically relevant embryological outcomes, providing a strong foundation to support and enhance clinical decision-making.

\end{document}